\documentclass[aps,floatfix,pra,superscriptaddress,reprint,showpacs,10pt,preprintnumbers,longbibliography]{revtex4-1}
\usepackage[utf8]{inputenc}
\usepackage[pdftex]{graphicx}
\usepackage{subfigure}
\usepackage{float}
\usepackage{amssymb}
\usepackage{amsmath}  
\usepackage{dsfont}
\usepackage{array}
\usepackage{bm}
\usepackage{mathrsfs}
\usepackage{pifont}
\usepackage{multirow}
\usepackage{upgreek}
\usepackage[dvipsnames]{xcolor}
\usepackage{soul}
\usepackage{verbatim}
\usepackage{slashed}
\usepackage[braket,qm]{qcircuit}
\usepackage{cleveref}
\usepackage{bm}
\usepackage{braket}
\usepackage{dcolumn}
\usepackage{tabularx}
\definecolor{mymagenta}{RGB}{200, 0, 100}
\definecolor{myblue}{RGB}{45, 48, 146}

\graphicspath{{figures/}}

\begin{document}

\title{Variational Quantum Eigensolver Approach to Prime Factorization on IBM's Noisy Intermediate Scale Quantum Computer}

\affiliation{Computation-Based Science and Technology Research Center, The Cyprus Institute, 20 Kavafi Street, 2121 Nicosia, Cyprus}
\affiliation{Deutsches Elektronen-Synchrotron DESY, Platanenallee 6, 15738 Zeuthen, Germany}
\affiliation{Northeastern University - London, Devon House, St Katharine Docks, London, E1W 1LP, United Kingdom}
\affiliation{Brandenburg University of Technology, Platz der Deutschen Einheit 1, 03046 Cottbus, Germany}

\author{Mona Sobhani}
\affiliation{Deutsches Elektronen-Synchrotron DESY, Platanenallee 6, 15738 Zeuthen, Germany}
\affiliation{Brandenburg University of Technology, Platz der Deutschen Einheit 1, 03046 Cottbus, Germany}

\author{Yahui Chai}
\affiliation{Deutsches Elektronen-Synchrotron DESY, Platanenallee 6, 15738 Zeuthen, Germany}
\email{yahui.chai@desy.de}

\author{Tobias Hartung}
\affiliation{Northeastern University - London, Devon House, St Katharine Docks, London, E1W 1LP, United Kingdom}

\author{Karl Jansen}
\affiliation{Computation-Based Science and Technology Research Center, The Cyprus Institute, 20 Kavafi Street, 2121 Nicosia, Cyprus}
\affiliation{Deutsches Elektronen-Synchrotron DESY, Platanenallee 6, 15738 Zeuthen, Germany}

\date{\today}
             
\begin{abstract}
This paper presents a hybrid quantum-classical approach to prime factorization. The proposed algorithm is based on the Variational Quantum Eigensolver (VQE), which employs a classical optimizer to find the ground state of a given Hamiltonian. A numerical study is presented, evaluating the performance of the proposed method across various instances on both IBM's real quantum computer and its classical simulator. The results demonstrate that the method is capable of successfully factorizing numbers up to 253 on a real quantum computer and up to 1048561 on a classical simulator. These findings show the potential of the approach for practical applications on near-term quantum computers.
\end{abstract}

\maketitle


\section{\label{sec:level1}Introduction}
The factorization of large integers is a computationally challenging task for classical computers. In 1994, Shor demonstrated that quantum computers have the potential for exponential speedup in solving the factorization problem \cite{shor}. However, due to current quantum hardware constraints, the largest number factored today using Shor's algorithm on a real quantum computer is 21~\cite{record26}.

While the large-scale implementation of Shor's algorithm in quantum computing remains a future challenge, near-term alternatives that harness the power of quantum computing to tackle optimization problems are being investigated. Through IBM's cloud-based quantum computer, Noise Intermediate-Scale Quantum (NISQ)~\cite{nisq} devices have enabled algorithm testing. It is important to note that these devices are not error-free and are vulnerable to hardware and mechanical quantum noise.

Variational Quantum Algorithms (VQAs) are particularly well-suited for NISQ devices. VQAs operate by establishing a parameterized quantum circuit, where the parameters are optimized using a classical computer to minimize a cost function that is decided by the state prepared by the quantum circuit and the target problem. This hybrid quantum-classical approach allows VQAs to effectively tackle problems that are difficult for purely classical or quantum methods. By maintaining relatively small quantum circuits and utilizing classical resources for optimization, VQAs can lead to significant results in finding the optimal solutions to realistic problems \cite{PhysRevE99013304, FGA_PRApllied, Amaro_2022}. In this paper, we focus on the Variational Quantum Eigensolver (VQE) \cite{vqe}, a specific type of VQA. The VQE is primarily designed to find the ground state energy of a quantum system. We propose to encode the prime factorization problem, mapped to an optimization problem as per Burges \cite{burges}, into a Hamiltonian. Our goal is to find the ground state, which corresponds to the solution of identifying the correct prime factors $p$ and $q$ for a given integer $n$, such that $n = p \times q$.

Investigating the potential of solving the prime factorization problem using a VQE contributes to our understanding of the practical applications of quantum computing and investigates the current limits of near-term quantum devices. The insights gained could lead to more efficient and effective solutions to complex computational problems in the future using a hybrid quantum-classical algorithm.

In this work, we aim to benchmark the performance of VQE combined with the Conditional Value at Risk (CVaR)~\cite{rockafellar2000cvar,Barkoutsos2020} of multiple different cost functions with respect to solving the prime factorization problem. However, in order to avoid any benchmarking effects of instance pre-processing, our benchmarks are performed without relying on any prior arithmetic simplifications. While this is not optimal for finding the largest possible values of $n$ that can be successfully factorized, we believe that it offers the fairest comparison of different quantum circuits and cost functions to be used in the CVaR-VQE. Nonetheless, it is important to note that we have successfully factorized 1,048,561 using 27 qubits in an ideal simulation and the number 253 with 9 qubits on a real quantum device. The paper is structured as follows: Section.~\ref{sec: related_work} provides a review of the relevant literature and presents the current factorization records using a hybrid quantum-classical approach. In Section.~\ref{sec: method}, we introduce our proposed method and discuss the theoretical background. Section.~\ref{sec: hardware_run} details the experimental setup, and Section.~\ref{sec: performance_factors} explores the factors that impact the algorithms' performance. Finally, we conclude our findings in Section~\ref{sec: conclusion}.

\section{Related Work}\label{sec: related_work}
This section presents the results of prime factorization records achieved through the use of hybrid quantum-classical algorithms. In 2001, Burges~\cite{burges} demonstrated that factoring a number \( n \) into its prime factors \( p \) and \( q \) can be mapped to an optimization problem as follows: 
\begin{equation}
    f(p, q) = (n - pq)^2.
    \label{eq:cost_function}
\end{equation} 
The minimum value of this function is achieved when the correct primes \( p \) and \( q \) are used. Various approaches have been proposed to find this minimum by using a Hamiltonian to find the ground state. Most methods use arithmetic simplifications of equations resulting from the binary multiplication table when multiplying $p$ and $q$ as binary numbers. For a detailed explanation of those approaches, we refer to \cite{record24}.

The adiabatic quantum computing model~\cite{Farhi_2001}, based on the adiabatic theorem~\cite{adiabatic}, has been used to factorize numbers using a gradually changing Hamiltonian. In 2008, the number 21 was factorized using 3 qubits on a  Nuclear Magnetic Resonance (NMR) machine~\cite{record3}, the record for factorization using this method is held by the number 291311, requiring 3 qubits~\cite{li2017highfidelity}. Furthermore, the number 35 was factorized using the adiabatic quantum computing model on a five-qubit IBM superconducting system~\cite{Saxena_2021}.

D-Wave Systems utilize the annealing method~\cite{dwave}  to solve the prime factorization problem, achieving a record by factorizing 376289 with 94 qubits~\cite{record13}. A comparison of Quadratic Unconstrained Binary Optimization (QUBO) and Higher-Order Unconstrained Binary Optimization (HUBO) models led to the factorization of 102454763 and 1000070001221 using 26 and 12 qubits respectively on a quantum simulator~ \cite{record22}.

The Variational Quantum Factoring (VQF) algorithm, introduced in 2018~\cite{record11}, uses the Quantum Approximate Optimization Algorithm (QAOA)~\cite{farhi2014quantum} to find the ground state of the corresponding Hamiltonian. In 2021, this method was used to factorize the numbers 1099551473989, 3127, and 6557 using 3, 4, and 5 qubits respectively~ \cite{record16}. The following year, a comparison of QAOA with the VQE led to the successful factorization of the numbers 25, 49, 91, and 247 using 6, 8, 10, and 13 qubits respectively, with VQE showing performance advantages~\cite{record19}. 

Other prime factorization approaches include the Quantum Imaginary Time Evolution (QITE) method~\cite{Motta_2019, McArdle_2019} to factorize numbers 1829 with 9 qubits using a real quantum computer~\cite{record17}, Grover's algorithm to factorize 1269636549803 using only 3 qubits~\cite{record27}. Additionally, several techniques combined allowed for the factorization of numbers using 3, 5, and 10 qubits~ \cite{record20}, and the Digitized Counteradiabatic Quantum Computing (DCQC) method was used to factorize 261980999226229 using 10 qubits ~\cite{record18}.

\section{Method}\label{sec: method}
The Variational Quantum Eigensolver (VQE) is a promising hybrid quantum-classical algorithm for NISQ devices, as outlined in \cite{vqe}. This algorithm utilizes a parametrized quantum circuit, which generates the quantum state $\ket{\psi(\bm{\theta})}$, where $\bm{\theta}$ denotes the set of tunable parameters. These parameters are optimized through a classical optimizer to minimize the energy expectation value, expressed as $\bra{\psi(\bm{\theta})} H \ket{\psi(\bm{\theta})}$.
This enables the VQE to search for the ground state of a Hamiltonian, which has the minimal energy value and encodes the optimal solution to our problem. 

Since the factoring problem is not directly associated with a Hamiltonian, there are many approaches one can pursue. Fundamentally, many possible choices come from the construction of a suitable Hamiltonian $H$ and the choice of parametrized quantum circuit that generates the state $\ket{\psi(\bm{\theta})}$. We will discuss the impact of the quantum circuit in more detail in Sec.~\ref{sec: performance_factors}. The focus of this current section will be on the Hamiltonian and associated cost function. 

The only fundamental constraint on the cost function and Hamiltonian comes from the fact that VQE attempts to minimize the value of the cost function. Hence, any Hamiltonian or cost function that encodes the solution to the factoring problem as a global minimum is viable. This freedom can be exploited through various means of pre-processing any given instance of the factoring problem in order to construct an optimized Hamiltonian that increases the chances of a successful factorization.

In contrast to many of the approaches presented in the previous section, we consider a generic approach that does not include simplifying equations arithmetically prior to the construction of the Hamiltonian. Instead, the cost function, as defined in Eq.~\ref{eq:cost_function}, is encoded directly into a Hamiltonian in order to prevent any effects arising from the pre-processing of instances. A direct VQE application then minimizes the energy expectation value to resolve the ground state of the constructed Hamiltonian. Once the VQE has identified the ground state, it can be concluded that the solution to the factoring problem has been found, with the values of $p$ and $q$ for $n=pq$ having been determined.

\begin{figure}[h!]
    \centering
    \includegraphics[width=1\linewidth]{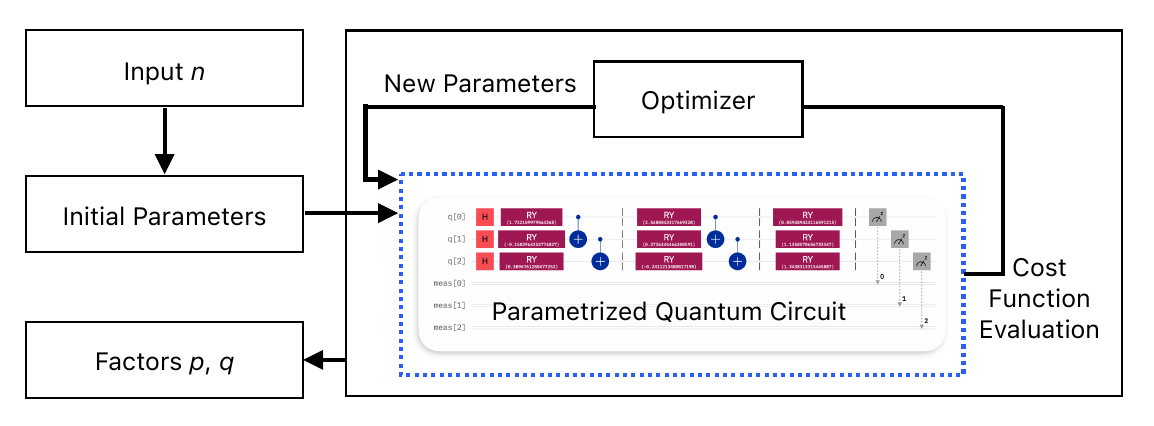}
    \caption{Steps of the VQE for prime factorization.}
    \label{fig:overview_final}
\end{figure}

This process is shown in Fig. \ref{fig:overview_final}. In Fig. \ref{fig:overview_final}, the ``Cost Function Evaluation'' step refers to measurements of the energy $\bra{\psi(\bm{\theta})} H \ket{\psi(\bm{\theta})}$ in the case of a direct VQE implementation. However, for combinatorial optimization problems such as the factoring problem, it is not necessarily advantageous to use the expected energy as a cost function. Since each individual shot result is a viable candidate for the ground state, the primary objective changes slightly from minimizing the energy to maximizing the probability of measuring the solution. While the two concepts are of course related, the latter offers more options. These include measuring functions of the Hamiltonian, such as $\lceil\log(\sqrt{H}+1)\rceil$ or $-\frac{1}{\sqrt{H}+\varepsilon}$, as well as using the Conditional Value at Risk (CVaR) which only considers the relatively low-energy shots for the cost function.

Below, we will explain the various steps shown in Fig.~\ref{fig:overview_final}. We begin by describing how to construct the Hamiltonian $H$ from Eq.~\ref{eq:cost_function}, the quantum circuit we employ, and the additional procedures used to evaluate and optimize the cost function.\\

\textbf{Step 1 - Construct Hamiltonian}: 
We use the cost function defined in Eq. \ref{eq:cost_function} where $n$ is a $B$-bit binary number
\begin{equation}
    n = \sum_{i=0}^{B-1}2^i n_i.
    \label{eq:n}
\end{equation}
We assume that $n$ is odd as otherwise the factorization problem is trivial. Similarly, the prime factors $p$ and $q$ can be represented by $N_p$-bit and $N_q$-bit binary numbers
\begin{equation}
    p = \sum_{j=0}^{N_p-1}p_j2^j \quad \text{and} \quad q = \sum_{k=0}^{N_q-1}q_k2^k.
    \label{eq:pq_bit}
\end{equation}
The bit-lengths of the integers $p$ and $q$ cannot be determined prior to the computation without further assumptions; however, the number of bits needed to represent $p$ and $q$ can be optimized. Assuming without loss of generality that $p \leq q$, it follows that $p \leq \sqrt{n}$. Consequently, the number of bits required to represent $p$ is $N_p = \lceil \log_2(\sqrt{n}) \rceil$. Additionally, if $n$ is not even, $q$ must be less than $\frac{n}{2}$, thereby reducing the number of bits for $q$ to $N_q = B-1$. Since $n$ is assumed to be an odd number, $p$ and $q$ are odd primes which implies that the least significant bits, denoted as $p_0$ and $q_0$, are always 1. These do not require encoding in qubits, leading to a total qubit requirement of $N = N_p+ N_q -2$.

Using this notation, the cost function can be written as: 
\begin{equation}
\begin{split}
    (n-pq)^2 &= n^2 - 2n \sum_{j=0}^{N_p-1} \sum_{k=0}^{N_q-1} p_jq_k 2^{j+k} \\
    &+ \sum_{j=0}^{N_p-1} \sum_{k=0}^{N_q-1}\sum_{l=0}^{N_p-1} \sum_{m=0}^{N_q-1} p_jq_kp_lq_m 2^{j+k+l+m}
\end{split}
\label{eq:cost_function_binary}
\end{equation}
Now, we encode the cost function into a Hamiltonian by substituting the binary variables $p_j, q_k$ (except $p_0$ and $q_0$) as defined in Eq. \ref{eq:pq_bit} with the Pauli Operator $\frac{I-Z}{2}$:
\begin{equation}
    \begin{aligned}
        p_j &\rightarrow \frac{I-Z_{p, j}}{2}, \ \text{with } j = 1, 2 \cdots N_p, \\
        q_k &\rightarrow \frac{I - Z_{q,k}}{2}, \ \text{with } k = 1, 2 \cdots N_q,
    \end{aligned}
\end{equation}
where $Z$ is the Pauli-Z operator. Thus we get the operator corresponding to factors $p, q$, and Hamiltonian
\begin{equation}
\label{eq:hamiltonian}
\begin{split}
    P &= 1+\sum_{j=1}^{N_p-1}2^{j-1}(I-Z_{p,j}),\\
    Q &= 1+\sum_{k=1}^{N_q-1}2^{k-1}(I-Z_{q,k}),\\
    H &= (nI-PQ)^2.
\end{split}
\end{equation}

We will use this Hamiltonian in our cost function evaluation part, explained in step 3.\\

\begin{figure}[h!]
    \centering
    \includegraphics[width=0.9\linewidth]{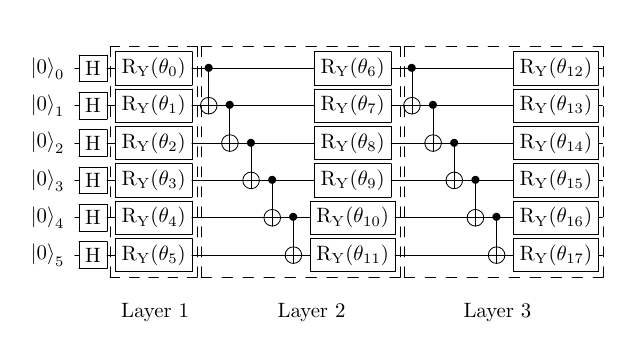}
    \caption{Linear-CNOT ansatz for $L=3$.}
    \label{fig:circuit_linear_cnot}
\end{figure}

\textbf{Step 2 - Quantum Circuit}: We use a hardware-efficient ansatz with linear-CNOT entanglement-layer for our method. For $N$ qubits and $L$ layers, we have $(N-1) \cdot (L-1)$ CNOT gates and $(N \cdot L)$ RY gates. The initial parameters of each RY gate are randomly chosen in the range of $(-\pi, \pi)$. An example circuit is shown in Fig.~\ref{fig:circuit_linear_cnot}.\\

\textbf{Step 3 - Cost Function Evaluation}:
For the cost function evaluation, we use the expectation value of the Hamiltonian $E=\langle \psi (\theta) |H| \psi(\theta)\rangle$ to determine how close the quantum state is to the solution. In more detail, we evaluate each quantum state $\ket{p,q}$ in the measurement output as follows:
\begin{equation}
    H\ket{p,q}= \lambda_{p,q} \ket{p,q} = (n - pq)^2 \ket{p,q}.
    \label{eq:ham_quantum_state}
\end{equation}
Here, the quantum state is defined as $\ket{p,q} = |p_{N_p-1} \cdots p_1\ q_{N_q-1} \cdots q_1\rangle$. For example, $\ket{p,q}=\ket{010001}$ with $N_q = 4, N_p = 4$ means $p$=0101 and $q$=0011, where $p_0 = q_0 = 1$ is added as the right-most bits of $p$ and $q$. Thus the bitstring $\ket{010001}$ corresponds to factors $p=5, q = 3$, and is the ground state of the corresponding Hamiltonian if $n=15$. The eigenvalue $\lambda_{p,q}$ of the state $|p, q\rangle$ is equivalent to our cost function value $(n - pq)^2$, with corresponding $n$, $p$, and $q$. This approach can be directly extended to expressions $f(\sqrt{H})$ of the Hamiltonian provided that $f(|n-pq|)$ can be efficiently evaluated. This will be the focus of Sec.~\ref{sec:cost-function}. Prior to Sec.~\ref{sec:cost-function}, we will only consider energy values $e_k=(n-pq)^2$ as obtained from single shots measuring $H$.

Instead of using the energy $E=\bra{\psi(\bm{\theta})} H \ket{\psi(\bm{\theta})}$ directly, the explicit cost function we evaluate is the Conditional Value at Risk (CVaR)~\cite{rockafellar2000cvar,Barkoutsos2020}. The procedure involves using $S$ shots to obtain $S$ measurement outcomes, resulting eigenvalues denoted as $E_S  = \{e_1, e_2, \cdots, e_S\}$. These eigenvalues are organized in ascending order and we consider the expectation of the lowest $\alpha$ percentile of the sorted list $E_S$ as the cost function evaluation
\begin{equation}
    \text{CVaR}(\alpha,S) = \frac{1}{\lceil \alpha S \rceil} \sum_{k=1}^{\lceil \alpha S \rceil} e_k. \\
    \label{eq:cvar}
\end{equation}
Thus, for $\alpha=1$, we recover the energy $E=\bra{\psi(\bm{\theta})} H \ket{\psi(\bm{\theta})}$, and in the limit $\lceil\alpha S\rceil=1$ we only use the shot with the lowest eigenvalue. The explicit choice of $\alpha$ therefore determines how much of the low-energy spectrum is used for the classical optimization. Intuitively, a smaller value of $\alpha$ is better since we want to maximize the chance of finding the solution to the factoring problem at least once while the high-energy contributions matter less. For example, if we wish to factorize $9$, the state $\ket{00}$ corresponds to $p=q=1$, $\ket{01}$ corresponds to $p=1$ and $q=3$, $\ket{10}$ corresponds to $p=3$ and $q=1$, and $\ket{11}$ corresponds to the solution $p=q=3$. We can consider two states of the quantum device, $\ket{\psi_1}=\sqrt{0.9}\ket{00}+\sqrt{0.1}\ket{11}$ which has energy $E=57.6$ and $\ket{\psi_2}=\sqrt{0.5}\ket{01}+\sqrt{0.5}\ket{10}$ which has energy $E=36$. Classical VQE would consider $\ket{\psi_2}$ better, while CVaR with $\alpha=0.1$ would prefer $\ket{\psi_1}$ with a score of $0.0$ over $\ket{\psi_2}$ with a score of $36$. For the factoring problem, $\ket{\psi_1}$ is the preferred state as it has a $90\%$ chance of observing the solution while $\ket{\psi_2}$ has a $0\%$ chance. This highlights how CVaR inherently improves the probability of observing low-energy states over simply optimizing the average energy. Since we only need to observe the correct factors $p$ and $q$ once, CVaR is the more natural choice of cost function. Choosing $\alpha$ as small as possible therefore seems to make sense, however it should be noted that the cost function landscape becomes discontinuous in the $\lceil\alpha S\rceil=1$ limit which hinders the classical optimization. Furthermore, the standard error of the estimated CVaR is unbounded in the $\alpha\to0$ limit for any given number of shots $S$. Thus, a compromise value of $\alpha$ needs to be chosen.

\textbf{Step 4 - Classical Optimization}: The CVaR value will then be used by the classical optimizer COBYLA \cite{Powell1994} to optimize the parameters $\theta$ for the RY gates in the quantum circuit. We refer to \cite{ragonneau2023pdfo} for an in-depth explanation of the COBYLA optimizer.

The COBYLA optimizer proves to be effective for solving optimization problems~\cite{D_ez_Valle_2021,ragonneau2023modelbased}. It is particularly useful when dealing with cost functions that are computationally expensive to evaluate. The classical optimizer optimizes the parameters $\theta$ until it has reached convergence or until it has reached the maximum number of iterations $50 \cdot N \cdot L$ set for the algorithm.

\section{Result of ideal simulation}\label{sec: ideal simulation}
This section shows the result simulated by IBM's classical simulator for quantum computers, the Aer simulator \cite{statevector}, which utilizes a statevector representation for the quantum state within the qiskit framework \cite{qiskit}. Table~\ref{table:highest_n} displays the values of the integer $n$ examined in this study. Each $n$ represents the largest number within a set of integers requiring the same number of qubits $N$.

\begin{table}[h!]
\centering
    \begin{tabular}{|c|c|c|c|}
    \hline
    $p$ & $q$ & $n$ & $N$ \\
    \hline
    3 & 5 & 15 & 3 \\
    \hline
    3 & 7 & 21 & 5 \\
    \hline
    3 & 19 & 57 & 6 \\
    \hline
    3 & 41 & 123 & 8 \\
    \hline
    11 & 23 & 253 & 9 \\
    \hline
    7 & 73 & 511 & 11 \\
    \hline
    3 & 337 & 1011 & 12 \\
    \hline
    23 & 89 & 2047 & 14 \\
    \hline
    61 & 67 & 4087 & 15 \\
    \hline
    19 & 431 & 8189 & 17 \\
    \hline
    11 & 1489 & 16379 & 18 \\
    \hline
    137 & 239 & 32743 & 20 \\
    \hline
    109 & 601 & 65509 & 21 \\
    \hline
    53 & 2473 & 131069 & 23 \\
    \hline
    349 & 751 & 262099 & 24 \\
    \hline
    269 & 1949 & 524281 & 26 \\
    \hline
    911 & 1151 & 1048561 & 27 \\
    \hline
    \end{tabular}
    \caption{The largest numbers $n$ for $N$ qubits. $p$ and $q$ are the prime factors of $n$, and $N$ is the number of qubits required to find $p$ and $q$.}
    \label{table:highest_n}
\end{table}

In our analysis of the algorithm, we focus on two key performance metrics: the success rate and the runtime. 
We define the success rate as the fraction of instances that have a fidelity, i.e. probability of the optimal solution, larger than the predefined fidelity threshold $t$ during the VQE process. Specifically, we monitor the fidelity during the VQE iteration. If the maximal fidelity throughout the VQE iterations is higher than the threshold $t$, we say that this VQE run is successful. For an instance achieving a fidelity threshold $t$, the probability of getting the solution at least once on a real quantum computer for $S$ shots is $P=1-(1-t)^S$. For example, when using 1000 shots and $t=0.01$, the probability of finding the solution at least once is $P=1-(1-0.01)^{1000}=99 \%$.

\begin{figure}[h!]
    \centering
    \includegraphics[width=1\linewidth]{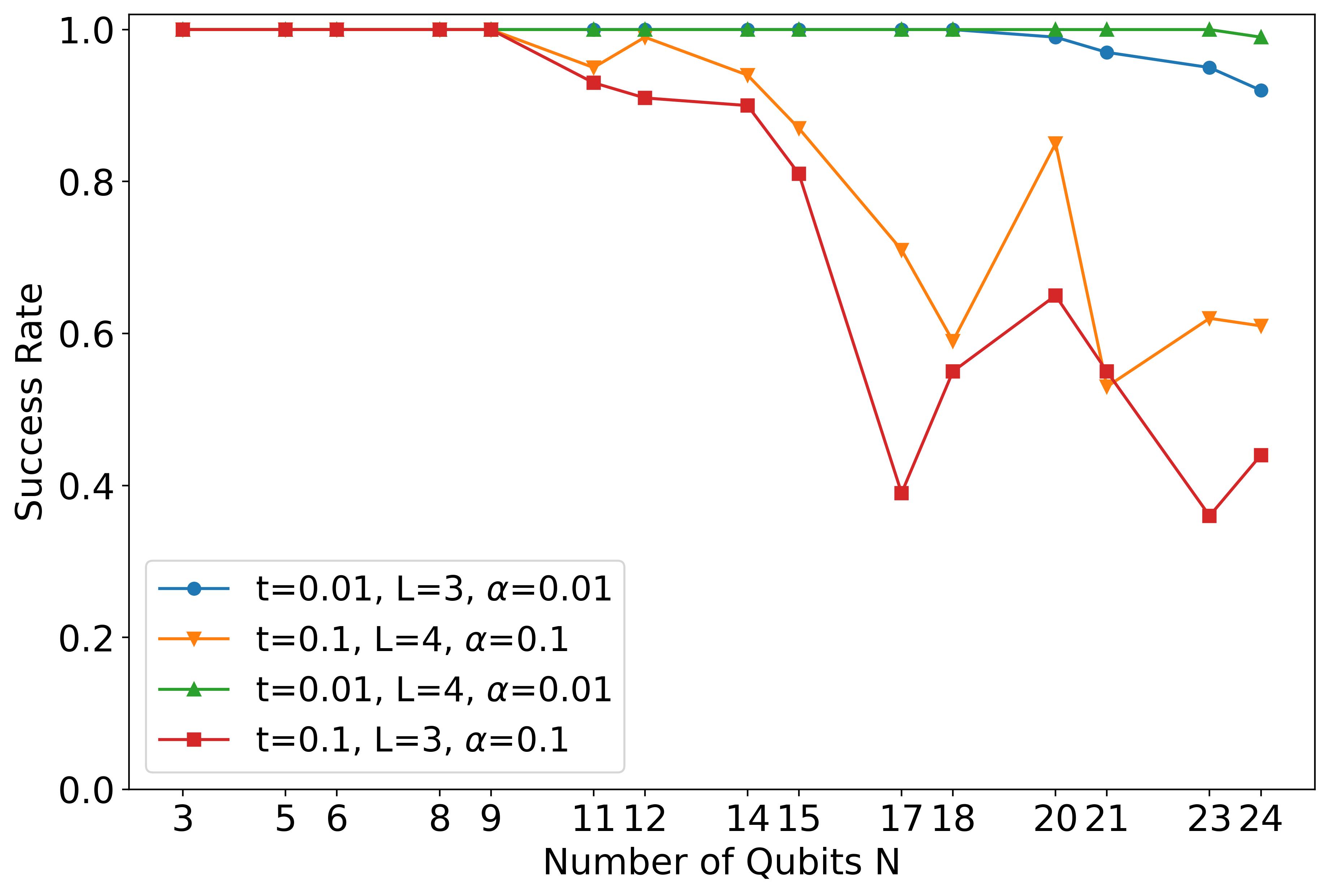}
    \caption{Success rate for largest numbers $n$ requiring $N$ qubits as outlined in Table.~\ref{table:highest_n}. }
    \label{fig:success_rate}
\end{figure}

Fig. \ref{fig:success_rate} depicts the success rate for varying numbers of qubits, layers $L$, CVaR percentages $\alpha$, and fidelity threshold $t$ set equal to $\alpha$. For each instance, the success rate is evaluated from 100 runs with random initial parameters. The result indicates $\alpha = 0.01$ achieves a higher success rate than $\alpha = 0.1$ in the ideal simulation. Additionally, increasing the number of layers enhances the performance for both CVaR values. Notably, the success rate almost remains $100\%$ across all problem sizes in the plot when $\alpha = 0.01$ and layer $L = 4$.

The solution for the highest number $n = 1048561$ factorized was obtained only with $t=0.01$, $L=3$, and $\alpha = 0.01$. Here, we ran 20 different initial parameters, but due to the computationally difficult calculations for $2^{27}$ numbers, we stopped the calculations after we found the solution.

\begin{figure}[h!]
    \centering
    \includegraphics[width=1\linewidth]{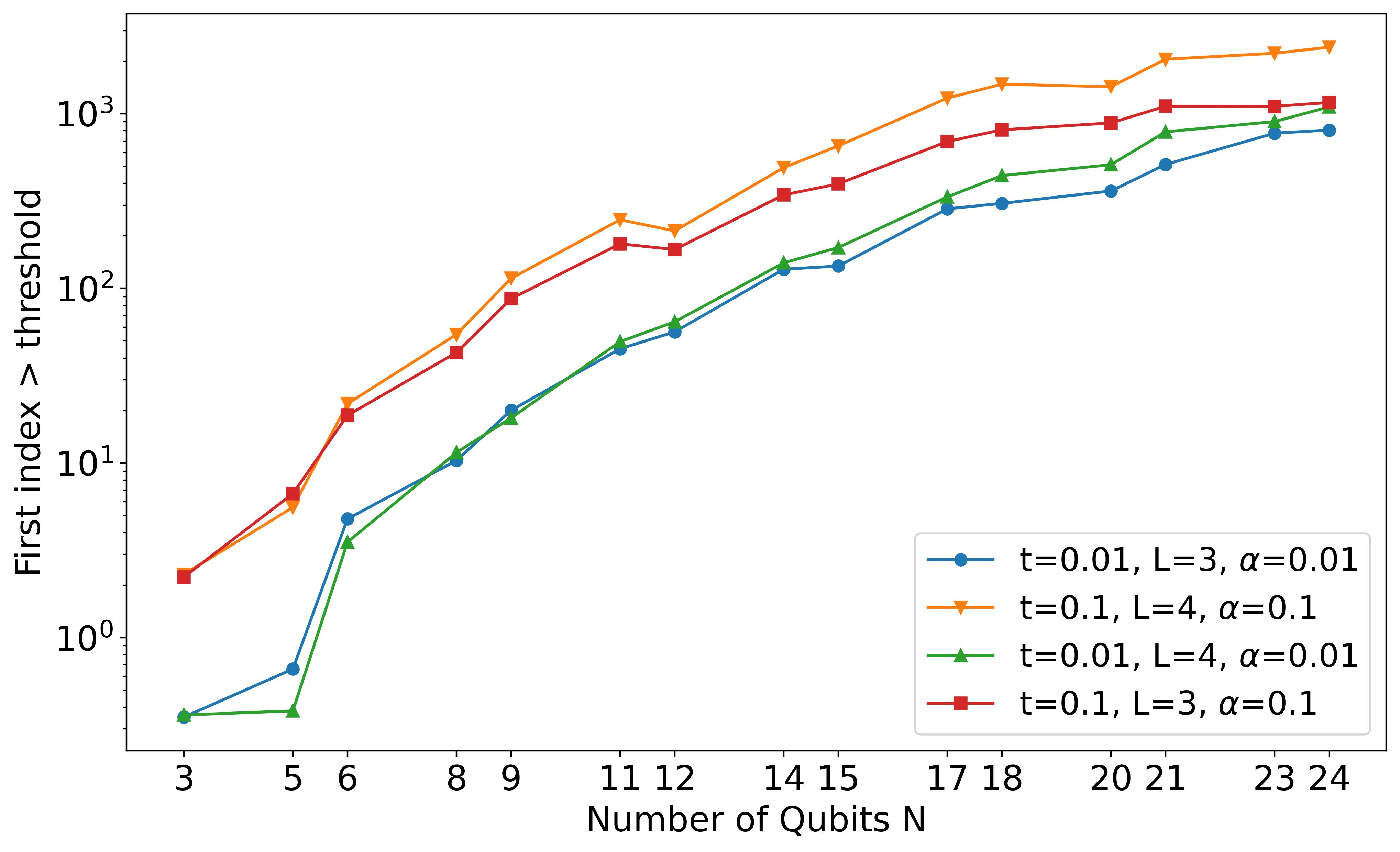}
    \caption{Scaling of minimal required iterations for different numbers $n$ requiring $N$ qubits.
 }
    \label{fig:plot_fidelity_avg}
\end{figure}

To assess the runtime of the VQE, we analyze the number of iterations required for convergence to the solution, which correlates with the number of function evaluations used by the COBYLA optimizer. We define the first iteration that is higher than the fidelity threshold as the minimal number of iterations required and compute the average of this metric across all successful runs. The scaling of these minimal necessary iterations is displayed in Fig.~\ref{fig:plot_fidelity_avg}. Notice that the CVaR coefficient $\alpha = 0.01$ not only results in a higher success rate but also achieves faster convergence.

\section{Real Hardware Results}\label{sec: hardware_run}
\begin{figure}[h!]
    \centering
    \includegraphics[width=0.9\linewidth]{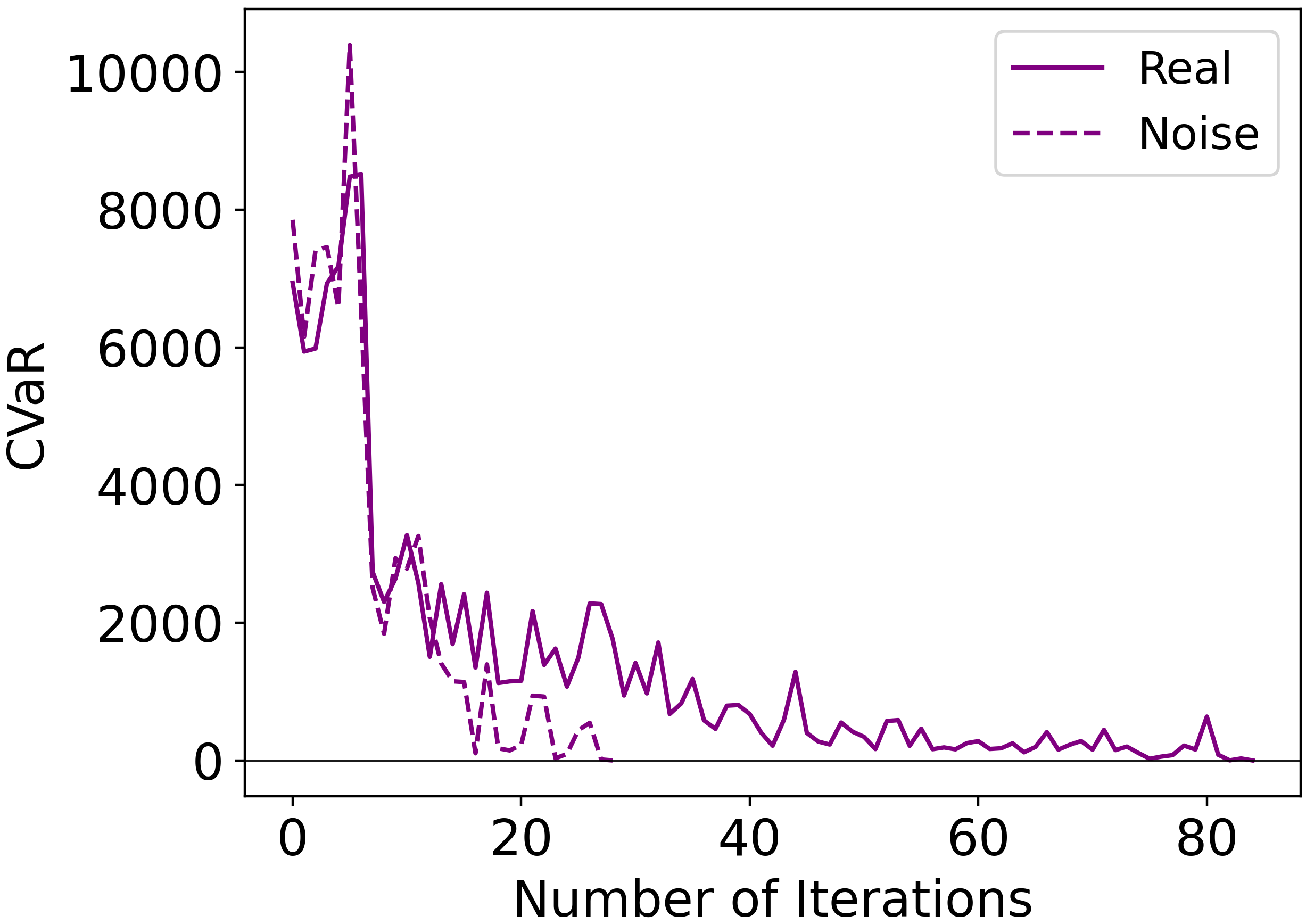}
    \caption{Hardware run for $n$=253, 3 layers, $\alpha=0.25$, 1000 shots requiring 9 qubits. The solid line indicates the hardware run result, and the dashed line represents the result from the noisy simulation with the same initial parameters.}
    \label{fig:cvar_real_noise_253}
\end{figure}

For the results obtained from the real quantum computer, a successful factorization is defined as the occurrence of the correct solution bitstring containing $p$ and $q$ in the measurement output. The integers 15, 21, 57, 123, and 253 were successfully factorized. Classical simulation with the noise model was employed using 1000 shots to identify the optimal initial parameters out of 350 distinct random initial parameters for the hardware run. Fig.~\ref{fig:cvar_real_noise_253} shows the cost function during the VQE process of number 253 from the hardware run, which converges to the optimal solution around 85 iterations.

The state vector simulation corresponding to infinite measurement shots in the last section indicates that the smaller alpha value $\alpha=0.01$ has a better performance. While considering the practical hardware run, only finite shots can be executed. As outlined in Ref.~\cite{Barkoutsos2020}, the standard error of the estimated CVaR($\alpha, S$) using $S$ shots scale as $\mathcal{O}(1/(S\alpha))$, implying that smaller $\alpha$ may introduce a larger uncertainty in estimating the cost function CVaR, which could mislead the classical optimizer during parameter updates. Therefore, it is crucial to strike a balance between the ideal performance and the statistical error when choosing the $\alpha$ value for the case of finite shots. In our hardware, we start from $\alpha=0.5$ and decrease it to $\alpha=0.25$ if $\alpha=0.5$ fails to find the solution. Tab.~\ref{tab:real_convergence} shows the first iteration index to converge to the minimal cost function value $0$. Note that for larger numbers $n=123, 253$, a smaller $\alpha$ value is used because $\alpha = 0.5$ was found too large to lead to convergence. This highlights the importance of selecting an appropriate $\alpha$ value in hardware run, prompting us to explore the impact of the $\alpha$ value in VQE more systematically in Sec.~\ref{sec: cvar_coefficient}.

\begin{table}[h!]
\centering
\begin{tabular}{>{\centering\arraybackslash}m{2cm} >{\centering\arraybackslash}m{2cm} >{\centering\arraybackslash}m{2cm} >{\centering\arraybackslash}m{2cm}}
Number $n$ & $N$ qubits & Req. iteration & $\alpha$\\
\hline 
15 & 3 & 4 & 0.5\\
21 & 5 & 9 & 0.5 \\
57 & 6 & 22 & 0.5 \\
123 & 8 & 76 & 0.25 \\
253 & 9 & 85 & 0.25\\
\hline
\end{tabular}
\caption{VQE iterations required to converge to the solution for a real hardware run.}
\label{tab:real_convergence}
\end{table}

\section{Exploring factors affecting VQE‘s performance}\label{sec: performance_factors}

\subsection{Difference between $p$ and $q$}
In this subsection, we investigate whether the difference between $p$ and $q$ has an impact on the success rate. The motivation behind this question comes from classical application cases. For example, in RSA applications, it is common to choose $p$ and $q$ to have the same bit-length. Thus, seeing a pattern in this range would be interesting. Similarly, if $p\approx q\approx \sqrt n$ or if one of the prime factors is very small, then brute-forcing a solution is very easy. Hence, it is interesting to see whether classically "easy" cases are easier for the quantum algorithm.

We conducted a comprehensive performance analysis for all numbers $n$ that require 20 qubits. For each of these numbers, we executed 50 random initial parameters and calculated the average success rate. It is important to note that we maintained consistency by using the same 50 initial parameters for each number $n$ across four different ansätze, as depicted in Appendix \ref{app:ansäzte} in Fig.~\ref{fig:anseatze_all}. It is important to emphasize that we choose the CVaR coefficient $\alpha=0.1$ here rather than the $\alpha=0.01$. The choice of $\alpha=0.01$ is able to achieve a success rate around $100\%$ for 20 qubits as shown in Fig.~\ref{fig:success_rate}, it would obscure the variance in success rates, which is the focus of this study. Hence, $\alpha=0.1$ was chosen to provide a more informative analysis by allowing us to observe variations in success rates more clearly, particularly with the difference between $p$ and $q$.

\begin{figure}[h!]
    \centering
    \centerline{(a) Linear-CNOT ansätze}
    \includegraphics[width=0.9\linewidth]{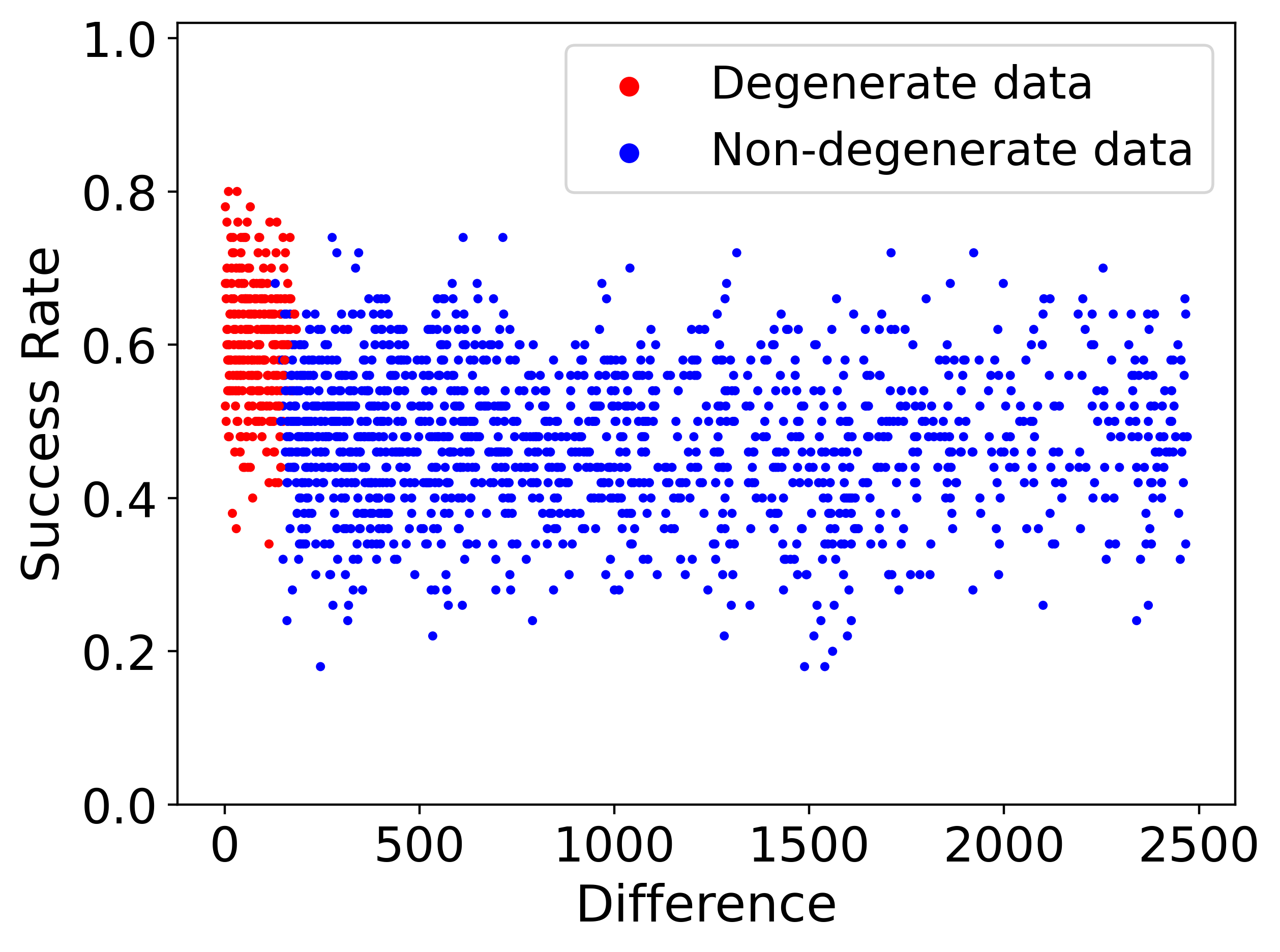}
    \centerline{(b) Parallel-CNOT ansätze}
    \includegraphics[width = 0.9\linewidth]{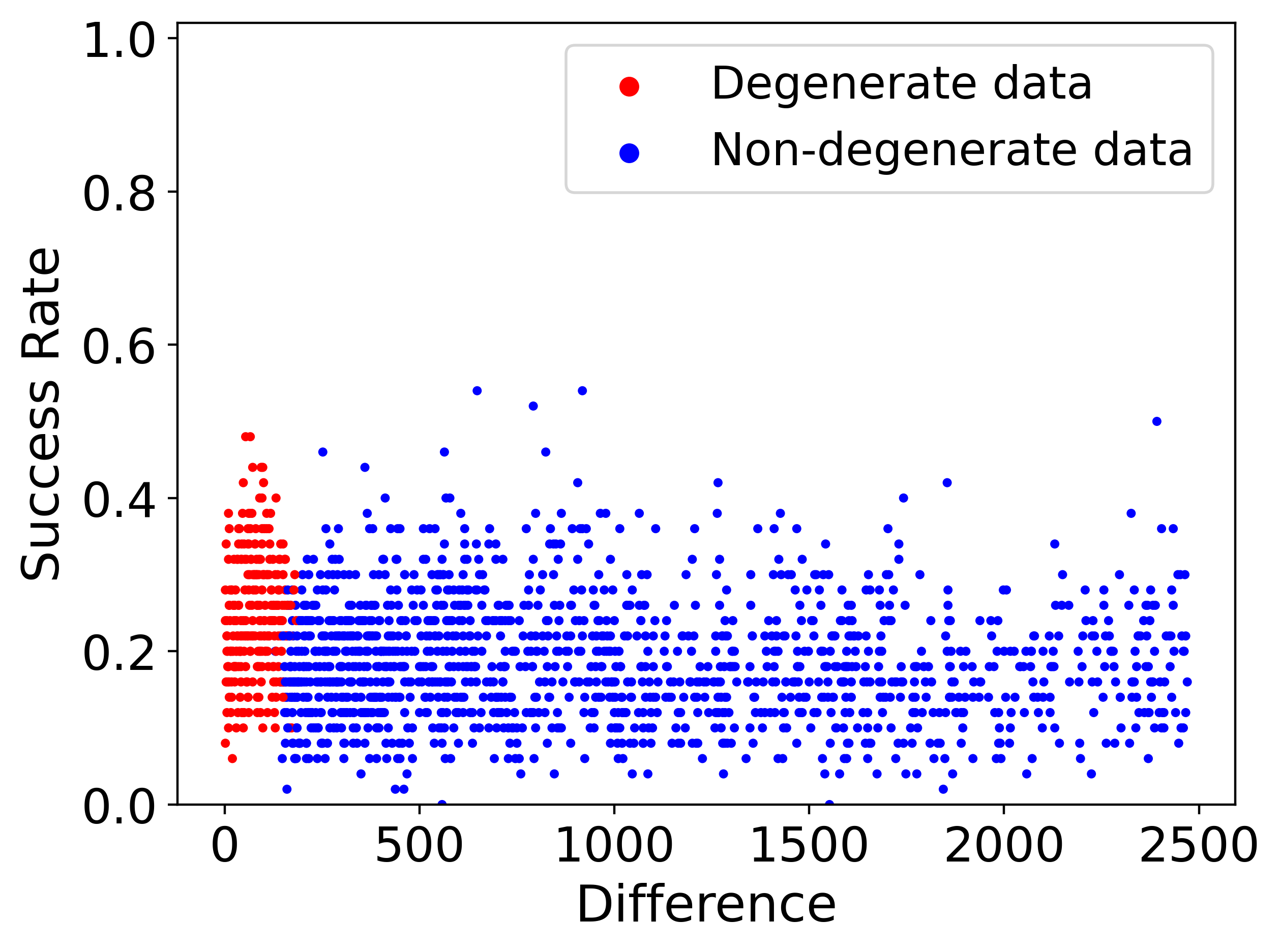}
    \caption{Average success rate for 50 random initial parameters for the ansatz (a) Linear-CNOT and (b) Parallel-CNOT for all possible numbers $n$ requiring 20 qubits, using $\alpha=0.1$, fidelity threshold $t=0.1$, 3 layers, and a maximum iteration number of 3000. The x-axis represents the absolute value of the difference between the prime factors of a specific $n$, i.e, $|p-q|$ for $p \cdot q = n$.}
    \label{fig:20_difference_success_rate}
\end{figure}

The results for two distinct ansätze are presented in Fig. \ref{fig:20_difference_success_rate} (a) for Linear-CNOT and in Fig. \ref{fig:20_difference_success_rate} (b) for Parallel-CNOT. The details of the circuit and results for two more ansatz Circular-CNOT and Parallel-CZ can be found in the appendix.~\ref{app:ansäzte}. The term ``degenerate data'' refers to the presence of two degenerate ground states in the associated Hamiltonian. This degeneracy typically occurs when the differences between the two prime factors are small, because the larger factor $q$ can also be accommodated by $N_p$ qubits in cases where $q \approx p \approx \sqrt{n}$. As a result, the success rate for the degenerate data is higher due to the doubled probability of identifying a solution, given the two potential bitstrings for the solution, either $\ket{p, q}$ or $\ket{q, p}$. Besides, Fig.~\ref{fig:20_difference_success_rate} shows Linear-CNOT has a much higher success rate than the Parallel-CNOT across all differences values, which indicates that the selection of the ansätze plays a critical role in the performance of the VQE in prime factorization problem. It is currently unknown what the primary factors influencing ansatz effectiveness are and thus how to find the optimal ansätze for the prime factorization is still an open question, which needs to be further investigated.

To ascertain whether larger or smaller differences between $p$ and $q$ influence the success rate, we employed the linear least-squares regression, as described in reference \cite{regression}. Our findings indicate that, for all ansätze, there is a slightly negative slope, suggesting that larger differences result in lower success rates than those with smaller differences between $p$ and $q$. For example, fitting a linear regression to the success rate of the Linear-CNOT ansatz shown in Fig.~\ref{fig:20_difference_success_rate}(a) yields that the success rate is approximately $0.5-3.7\cdot10^{-5}|p-q|$. While this slope of $-3.7\cdot10^{-5}$ is very small, its standard deviation is $3.5\cdot10^{-6}$, i.e., there is a $10$-standard deviation separation of the slope from $0$, making it highly significant. Similarly, for the Parallel-CNOT case shown in Fig.~\ref{fig:20_difference_success_rate}(b), the slope is negative with a $7.3$-standard deviation separation from $0$. However, when examining the linear regression of the degenerate and non-degenerate data separately, the slopes are generally less than 3 standard deviations away from $0$. Furthermore, we observe that the slope for the ansatz Parallel-CNOT is slightly positive ($1.4$ standard deviations positive), indicating that larger differences between $p$ and $q$ have a higher success rate. A notable exception is the case of a Parallel-CZ ansatz, which is identical to Parallel-CNOT except each CNOT-gate is replaced with the corresponding CZ-gate, for which the degenerate data shows a $5$-standard deviation positive slope. Consequently, at this time, it appears as if degeneracy slightly improves the success rate but no definitive conclusion can be made about the success rate as a function of $|p-q|$ beyond any potential degeneracy effects. Instead it seems more prudent to expect that the small variations in perceived difficulty are instance and ansatz dependent.

\subsection{CVaR coefficient}\label{sec: cvar_coefficient}

Analyzing our initial hardware runs, it seems that smaller values of $\alpha$ are required as the system size increases. Furthermore, based on the ideal simulation results with infinite shots, as shown in Fig.~\ref{fig:success_rate}, it is evident that a smaller $\alpha$ leads to a higher success rate. However, when using a finite number of shots $S$, the native intuition is that the standard error of estimated CVaR scales as $\mathcal{O}(1/(S\alpha))$, indicating a smaller $\alpha$ might result in fewer effective measurements and larger standard error, which may impact the optimization process of VQE. Therefore, it is crucial to balance the improved performance of a smaller $\alpha$ with the associated statistical error. In this subsection, We explore the behavior of the standard error for different $\alpha$ with finite shots.

The standard error with a given number of shots $S$ is defined as:
\begin{equation}
    \sigma = \sqrt{\frac{\sum_{k=1}^{\lceil \alpha \cdot S \rceil } \left(e_k - \text{CVaR}(\alpha,S)\right)^2}{\lceil \alpha \cdot S \rceil \left(\lceil \alpha \cdot S \rceil - 1 \right)}},
    \label{eq: standard error}
\end{equation}
where the measured eigenvalues are sorted in ascending order $e_1 \leq e_2 \leq \cdots \leq e_S$, and  CVaR($\alpha, S$) is detailed in Eq.~\eqref{eq:cvar}.
As shown in the above equation, there are two factors that impact the value of the standard error: firstly,
a smaller $\alpha$ leads to less effective measurements $\lceil \alpha \cdot S \rceil$, which will lead to a smaller denominator in Eq.~\ref{eq: standard error}; on the other hand, smaller $\alpha$ means selecting a smaller portion of the left-tail of the measured results, which will lead to a smaller numerator in Eq.~\ref{eq: standard error}, these two factors will compete with each other.

\begin{figure}[h!]
    \centering
    \centerline{(a) 1000 shots}
    \includegraphics[width=0.9\linewidth]{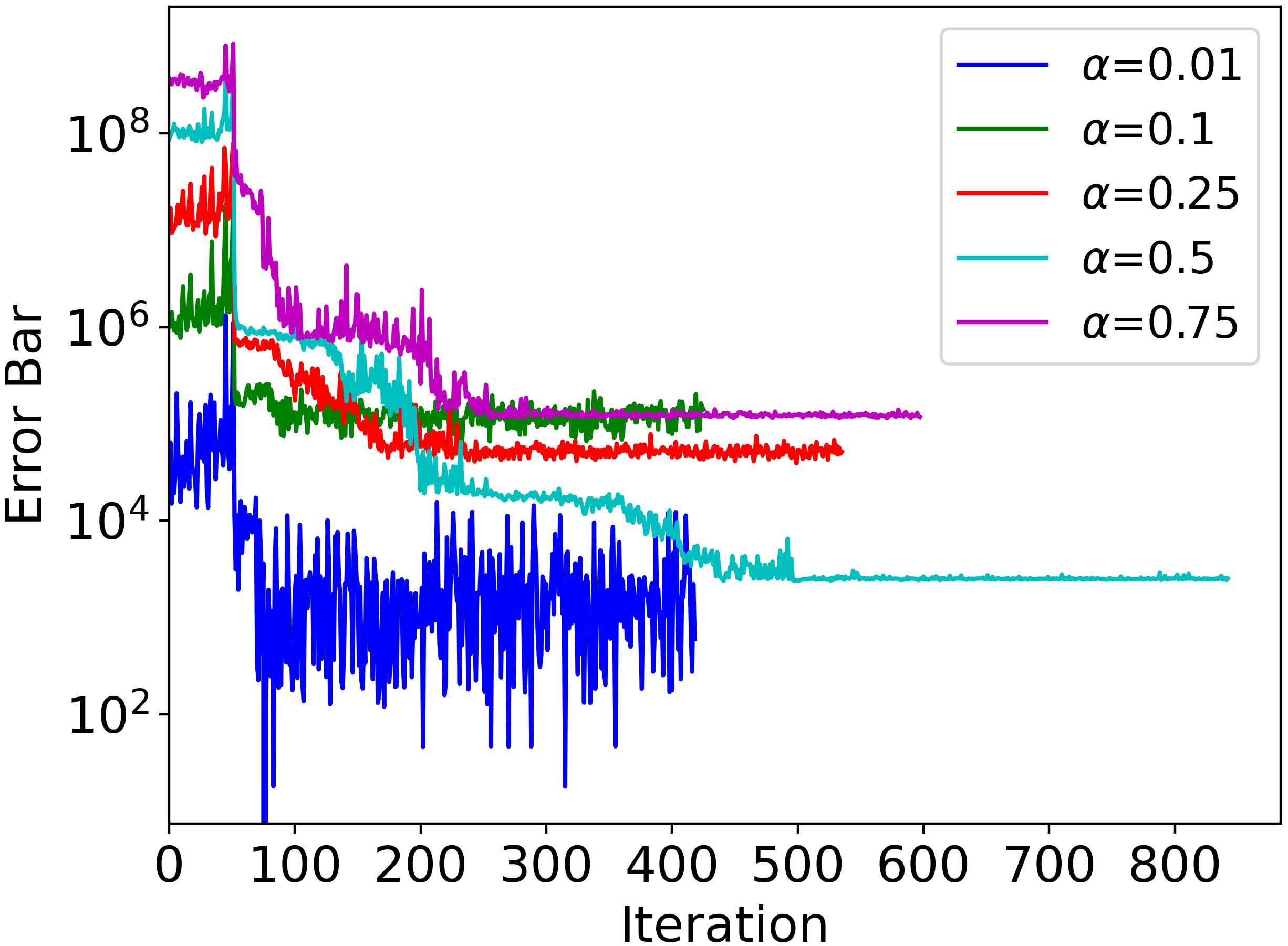}
    \centerline{(b) 10000 shots}
    \includegraphics[width = 0.9\linewidth]{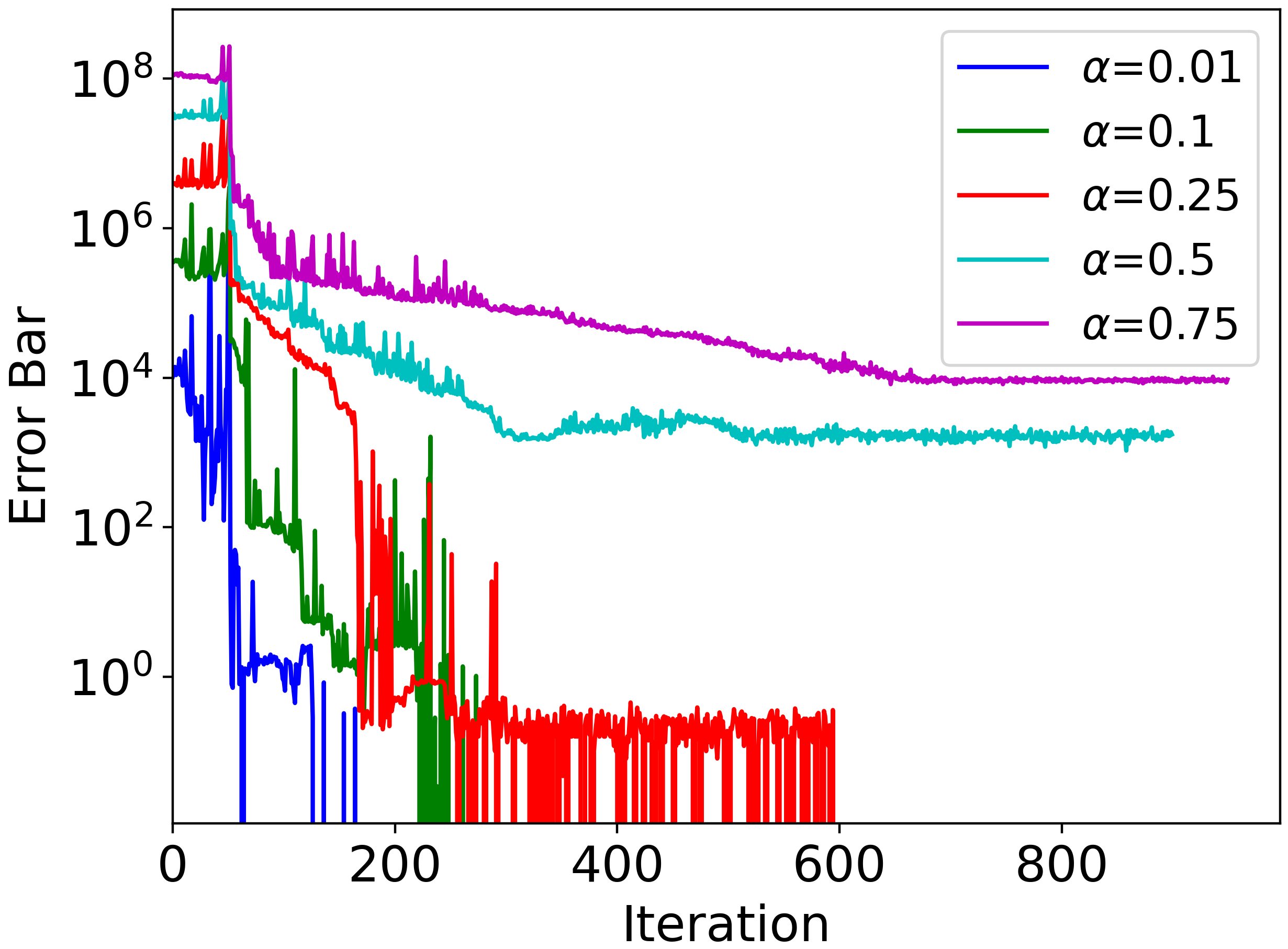}
    \caption{Error bar in VQE process for different $\alpha$ values for $n$=8189, requiring 17 qubits. }
    \label{fig:error_bar}
\end{figure}

Fig.~\ref{fig:error_bar} displays the standard error of CVaR for different $\alpha$ values in the VQE process used to factorize $n=8189$, requiring $17$ qubits. For consistency, VQE for all the $\alpha$ values starts at the same initial state $\ket{+}^{\otimes N}$. At the beginning of the VQE process, smaller $\alpha$ results in a smaller standard error for both $1000$ shots and $10000$ shots, which indicates the numerator in Eq.~\ref{eq: standard error} dominate. As optimization progresses, the VQE with different $\alpha$ might converge to different local minima, potentially leading to a smaller standard error for larger $\alpha$, as depicted in Fig.~\ref{fig:error_bar}(a). While Fig.~\ref{fig:error_bar}(b) with a larger shot number $S=10000$ indicates a clear trend that smaller $\alpha$ is more likely to have a smaller standard error during the VQE process. Nonetheless, the $\alpha = 0.01$ always keeps the smallest standard error during the whole VQE process for both shots $S=1000$ and $S=10000$. 

\begin{figure}[h!]
    \centering
    \centerline{(a) 1000 shots}
    \includegraphics[width=0.9\linewidth]{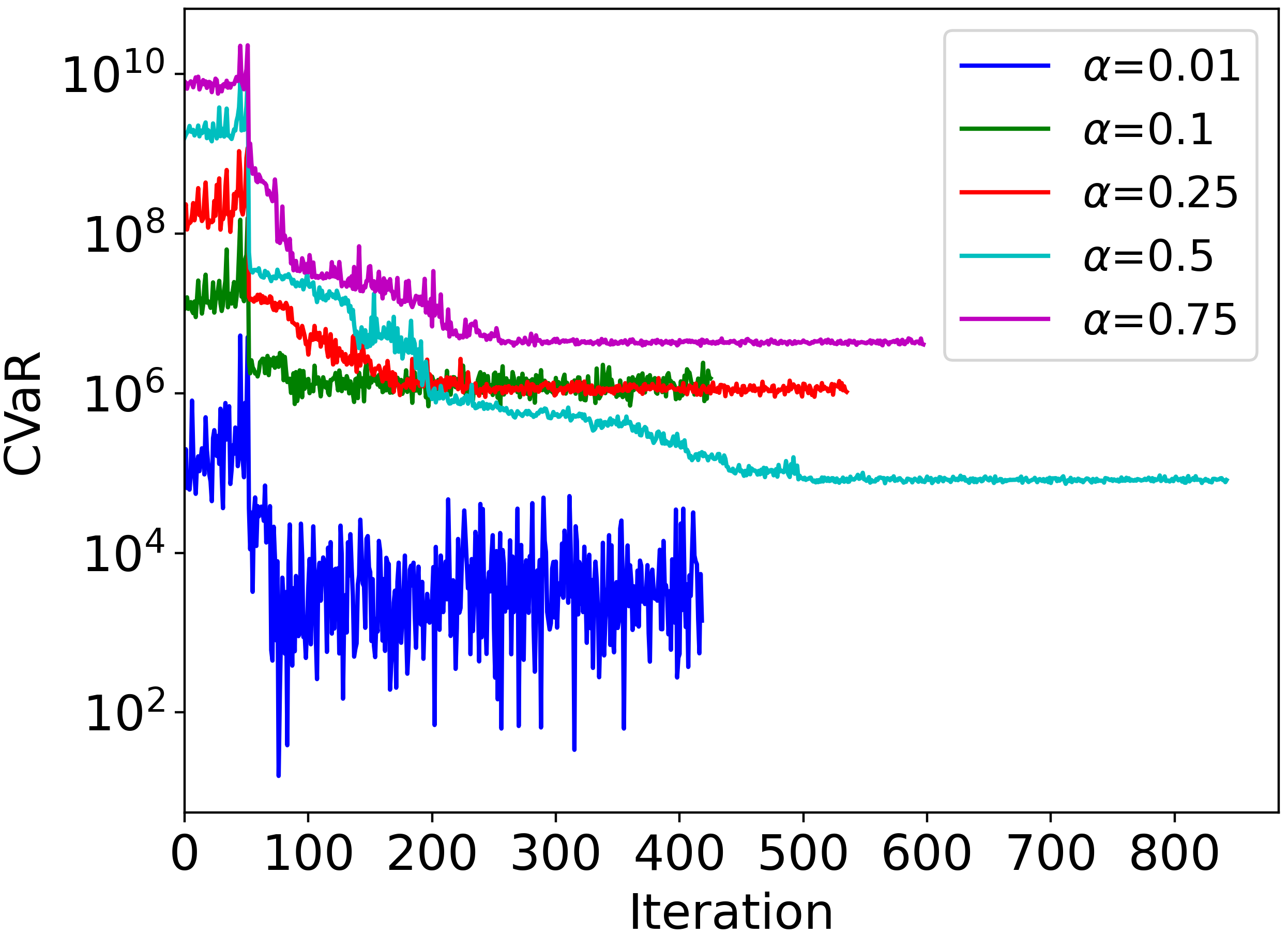}
    \centerline{(b) 10000 shots}
    \includegraphics[width = 0.9\linewidth]{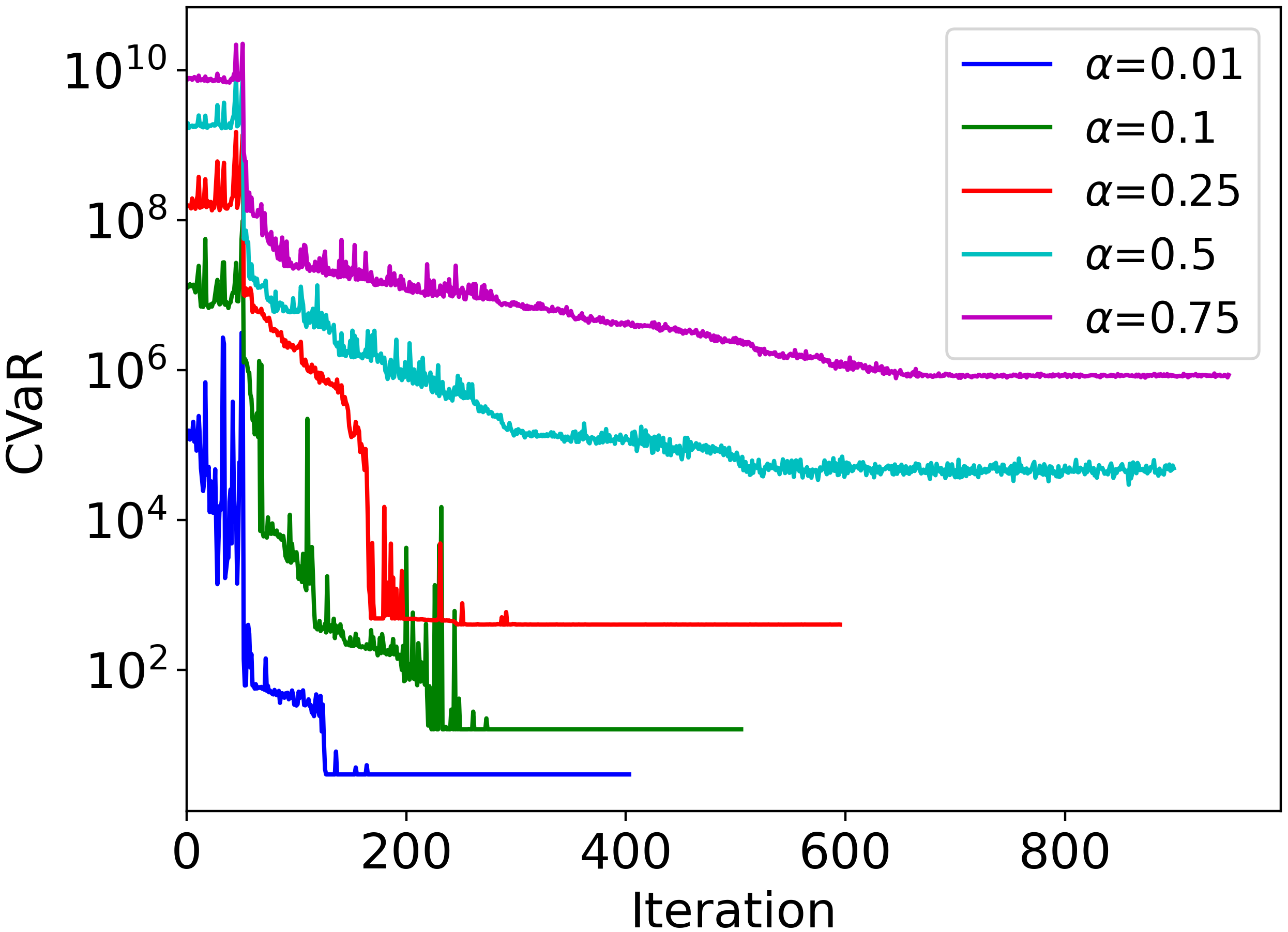}
    \caption{CVaR in VQE process for different $\alpha$ values for $n$=8189, requiring 17 qubits.}
    \label{fig:cvar}
\end{figure}

Furthermore, the corresponding cost function value using different $\alpha$ in the VQE process is shown in Fig.~\ref{fig:cvar}. With shots $S=1000$, the smaller $\alpha$ can not certainly result in a better result, which might due to the optimizer being misled by the large standard error; when increasing the shots number to reduce the standard error, smaller $\alpha$ can keep smaller standard error and result in better performance.

In summary, smaller $\alpha$ does not necessarily lead larger standard error in CVaR when using finite shots. Although only $\alpha$ portion of sampled result is selected to do the average, which reduce the dominator in Eq.~\ref{eq: standard error}, a more concentrated distribution with a smaller $\alpha$ also results in a smaller numerator, ultimately leading to a smaller standard error of CVaR. Additionally, although smaller $\alpha$ achieves smaller standard error in the beginning, a relatively large statistic error might still mislead the optimization and result into a worse local minimum, while with a sufficiently large number of shots, smaller $\alpha$ can keep a small standard error and good stable performance. Our result indicate the most stable choice of CVaR value is $\alpha=0.01$ among our testing values $\alpha \in \{0.01, 0.1, 0.25, 0.5, 0.75\}$. This is in agreement with our initial observation that smaller $\alpha$ is better. However, it should be noted that this can only be expected to be true within reasonably large numbers of shots used. More precisely, if $\alpha\le\frac1S$, then the CVaR will only use the smallest observed value which leads to discontinuous optimization landscapes. Thus, $\alpha$ should be kept significantly larger then $\frac1S$, but within that regime a smaller $\alpha$ seems to be better. Finding the critical point at which $\alpha$ becomes too small is likely highly problem and instance specific, so we recommend users to check for their individual applications.

\subsection{Cost functions}\label{sec:cost-function}

In section~\ref{sec: ideal simulation}, we present the success rate of the CVaR-VQE in the ideal simulation with infinite measurement shots, the success rate is almost $100\%$ if using CVaR coefficient $\alpha = 0.01$, as shown in Fig.~\ref{fig:success_rate}. This indicates the ability of CVaR-VQE to find the optimal solution when the expectation of cost function can be estimated with perfect accuracy. While in practical scenarios, only finite shots can be used when estimating the expectation of cost function, the statistic error as described in Eq.~\eqref{eq: standard error} may mislead the optimizer and disrupt the optimization process to the optimal solution. In this subsection, we investigate the effect of the finite shots on success rate. Furthermore, we explore different cost function types, trying to improve the success rate in the case of finite shots.

\begin{figure}
    \centering
    \includegraphics[width=0.9\linewidth]{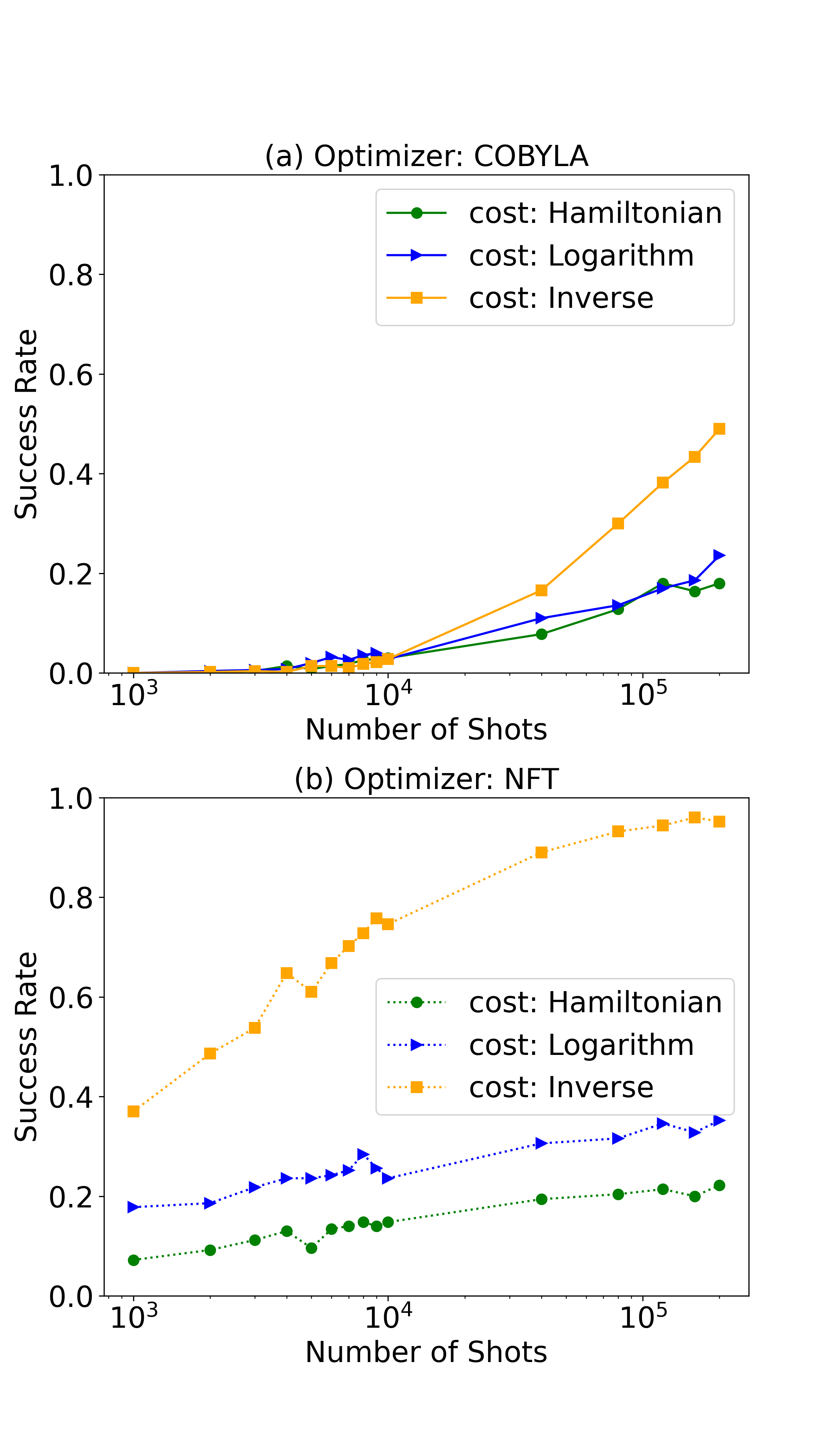}
    \caption{Success rates a function of the number of measurement shots for different types of cost functions using the COBYLA optimizer (a) and NFT (b).}
    \label{fig: sr_costs}
\end{figure}

Using $n=32743$ as an example, which requires 20 qubits, we explore the success rate with finite shots among 500 runs with random initial parameters, the number of measurements for estimating the CVaR value ranges from 1000 to 200000. The ansatz layer $L=3$ and CVaR coefficient is selected as $\alpha=0.01$, which also has the best performance with finite shots as indicated in the last subsection. Besides the optimizer COBYLA we used in the previous section,  we also test the performance of NFT~\cite{nft} in the case of finite shots. As shown in Fig.~\ref{fig: sr_costs}(a), using the Hamiltonian cost function Eq.~\eqref{eq:hamiltonian}, the success rate is almost zero when the number of shots is smaller than 10000 when using COBYLA, NFT performs better in this scenario. However, for both optimizers, COBYLA and NFT, the success rates are only around $20\%$ for even 200000 shots.

The Hamiltonian cost function can produce extremely large eigenvalues. For instance, in the case $p\sim q\sim 1$, the function $(n-pq)^2 \sim n^2 \sim 10^9$, which means optimizing the low-energy part needs very high accuracy when estimating the expectation value. For example, the change from the suboptimal solution $(n-pq)^2=1$ to the optimal solution $(n-pq)^2=0$  might only result in a relative change in the CVaR value of approximately $\frac{1}{n^2} \sim 10^{-9}$, which might even smaller than the standard error of CVaR in the case of finite shots. Therefore, we consider additional cost functions. One option is a logarithmic cost function which compresses the high-energy domain, making it less sensitive to large eigenvalues. Another option is an inverse type of cost function which not only compresses the high-energy domain but also introduces a large energy gap between the solution and suboptimal solutions. The following are the cost functions used:

\begin{enumerate}
    \item \textit{Hamiltonian}: $(n-pq)^2$,
    \item \textit{Logarithm}: $\lceil \log(|n - pq| + 1) \rceil$,
    \item \textit{Inverse}: $-\frac{1}{|n - pq| + 0.001}$.
\end{enumerate}

\begin{figure}
    \centering
    \includegraphics[width=0.95\linewidth]{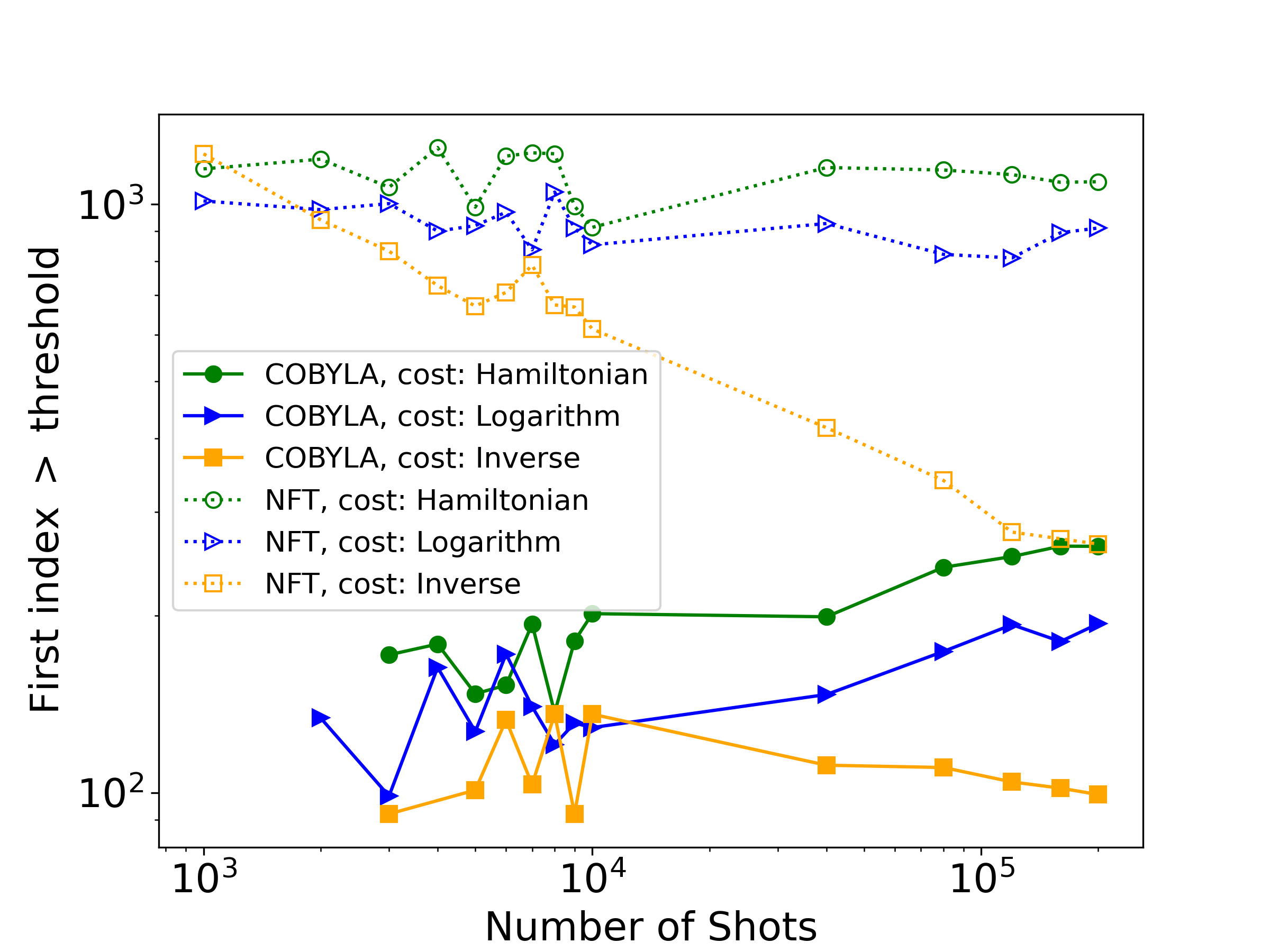}
    \caption{Number of iterations required to reach the fidelity threshold for the successful runs, with various types of cost functions and using COBYLA or NFT optimizers.}
    \label{fig:iter_cost}
\end{figure}

Since each observed bitstring encodes the binary representation of a candidate solution for p and q, we can directly compute the cost function values from each shot. As shown in Fig.~\ref{fig: sr_costs}, the logarithmic cost function has comparable performance with the Hamiltonian cost function, while the inverse type of cost function has much better performance, especially with the NFT optimizer. However, it is important to note that the NFT optimizer needs a larger number of function evaluations to achieve the fidelity $1\%$ as shown in Fig.~\ref{fig:iter_cost}. It is not clear at the moment whether the logarithmic and inverse type of cost function are related to global operators which lead to the problem of barren plateau~\cite{Cerezo_2021}. To address this, future work will focus on identifying suitable initial parameters for the inverse cost function to overcome the potential barren plateau issue and reduce optimization time. This will be a critical area of exploration moving forward.

\section{Conclusion}\label{sec: conclusion}

In this work, we benchmark the performance of CVaR-VQE in solving the prime factorization problem without relying on prior arithmetic simplifications. The number $1048561$ requiring $27$ qubits is factorized in the ideal simulation, and the number $253$ requiring $9$ qubits is factorized by the IBM hardware. 

Furthermore, we explore various factors influencing VQE's performance on the prime factorization problem. Firstly, we observe the difference between two factors $|p-q|$, which is crucial in the difficulty of classical algorithms, does not impact the success rate of VQE a lot, while the choice of ansatz impacts VQE's performance on the factoring problem obviously. Secondly, we explore how the statistic error behaves across different CVaR coefficients $\alpha$. In contrast to naive intuition, the smaller $\alpha$ achieves a smaller standard error and better performance with a sufficient number of measurement shots. Especially, $\alpha=0.01$ performs best in our problem with finite shots. 

Finally, we use $\alpha=0.01$ to examine the scaling of success rate with shots number, the result indicates the Hamiltonian cost function needs a huge number of shots to improve the success rate, prompting us to explore a better cost function for VQE. We find an inverse-type cost function with the NFT optimizer can achieve a much higher success rate. However, the inverse-type cost function needs more iterations to converge, and it is an open question whether this cost function corresponds to a global observable and might have the problem of barren plateau. A possible way to overcome the above problems is to find a good initial parameter of the VQE for this inverse cost function, which shows significant improvement in reducing iteration number and mitigating the barren plateau problem in the QUBO problem~\cite{chai2024structureinspiredansatzwarmstart}. While it is nontrivial to extend this method to the inverse type cost function, we leave it to future work.

\begin{acknowledgments}

 This work is funded by the European Union’s Horizon Europe research and innovation funding programme under the ERA Chair scheme with grant agreement No. 101087126 and under the European Research Council (ERC) with grant agreement No. 787331. The Ministry of Science, Research and Culture of the State of Brandenburg within the Centre for Quantum Technologies and Applications (CQTA).  This work is supported with funds from the Ministry of Science, Research and Culture of the State of Brandenburg within the Centre for Quantum Technologies and Applications (CQTA).
 \begin{center}
        \includegraphics[width = 0.08\textwidth]{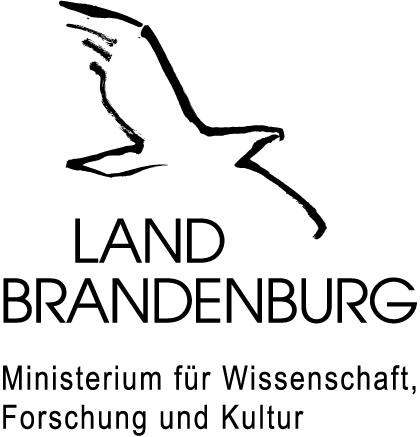}
\end{center}
\end{acknowledgments}

\begin{widetext}
\appendix
\section{Performance of different Ansätze with difference between $p$ and $q$}
\label{app:ansäzte}
Fig.~\ref{fig:anseatze_all} shows all four different ansätze that we tested. And Fig.~\ref{fig: two_more_ansatz} shows the performance of Parallel-CZ and Circular-CNOT ansätze regarding the difference between two factors $p$ and $q$.

\begin{figure*}[h!]
    \centering
    \includegraphics[width=1\linewidth]{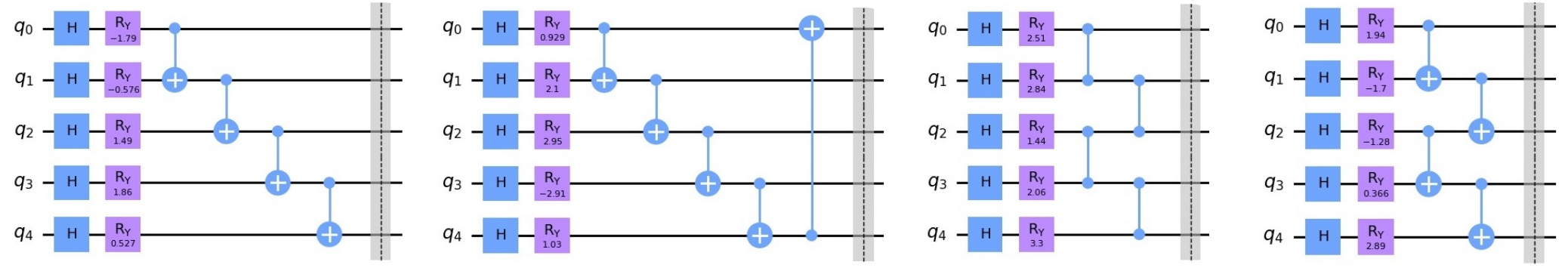}
    \caption{Ansätze from left to right: Linear-CNOT, Circular-CNOT, CZ-parallel, CNOT-parallel. Circuits illustrated with Qiskit library~\cite{qiskit}.\\
    }
    \label{fig:anseatze_all}
\end{figure*}

\begin{figure}[h!]
    \centering
    \subfigure[\ Parallel-CZ ansätze]{
    \includegraphics[width=0.45\linewidth]{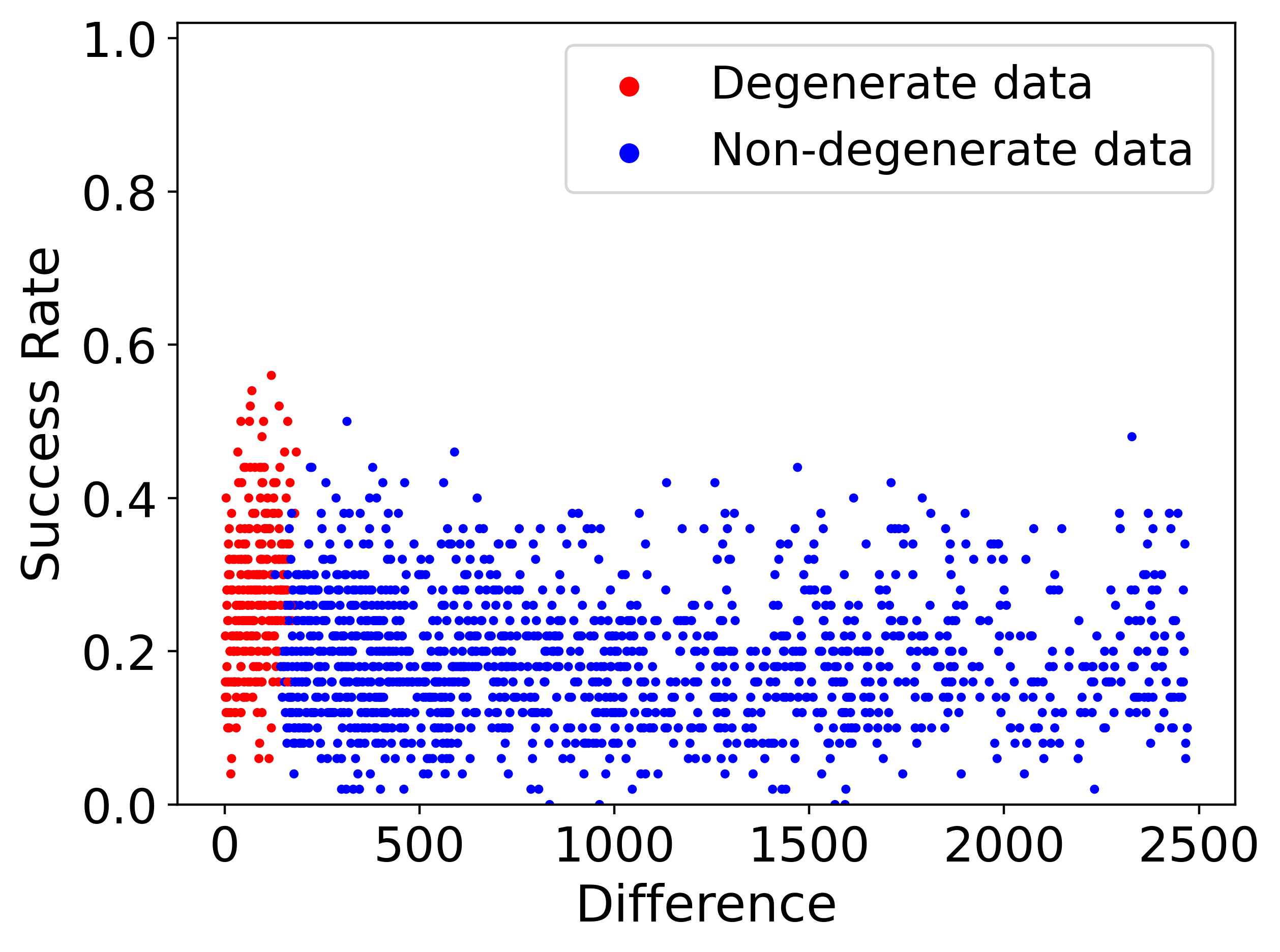}}
    \subfigure[\ Circular-CNOT ansätze]{
    \includegraphics[width = 0.45\linewidth]{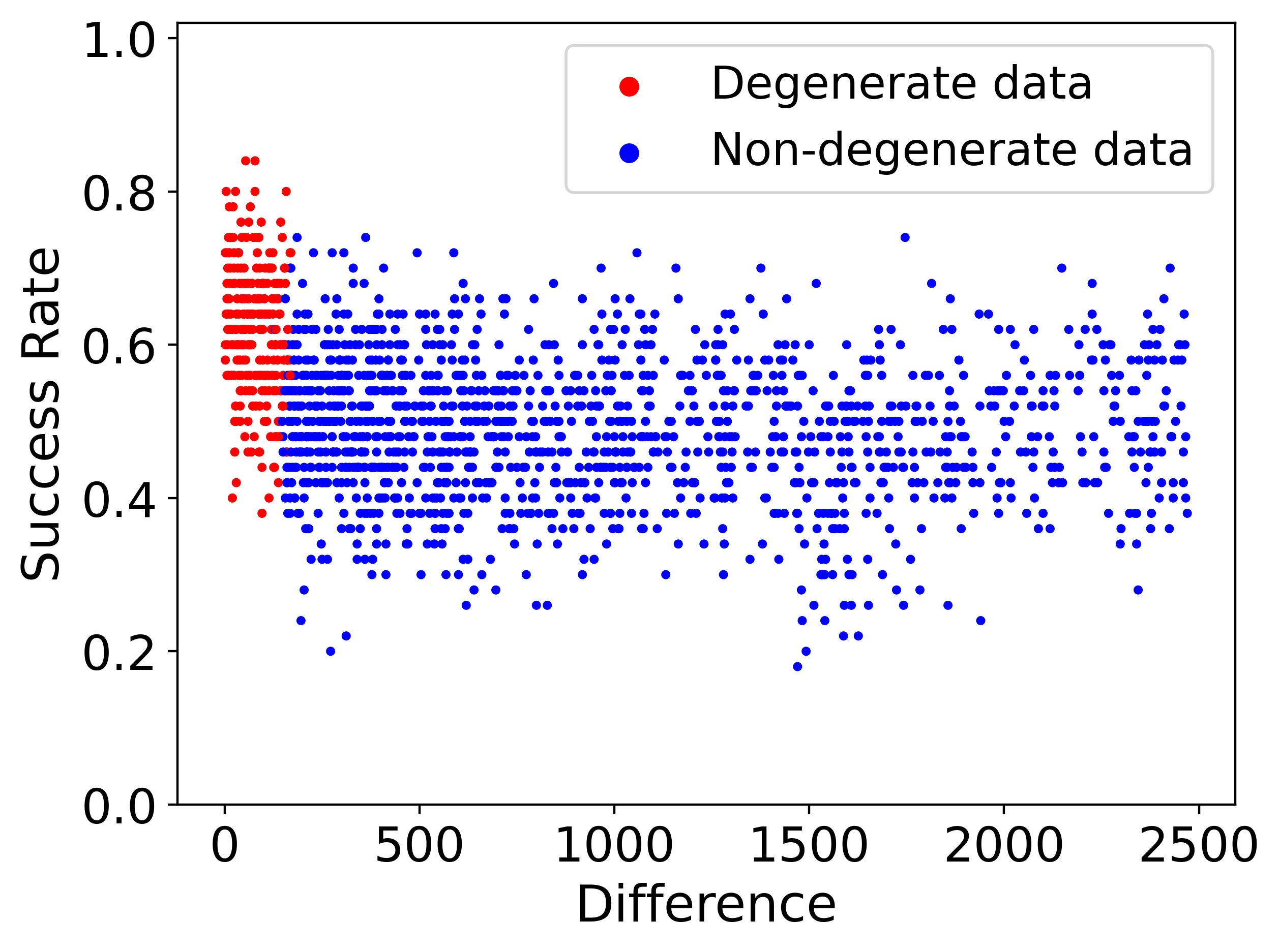}}
    \caption{Average success rate for 50 random initial parameters for the ansatz (a) Parallel-CZ and (b) Circular-CNOT ansätze for all possible numbers $n$ requiring 20 qubits, using $\alpha=0.1$, fidelity threshold $t=0.1$, 3 layers, and a maximum iteration number of 3000. The x-axis represents the absolute value of the difference between the prime factors of a specific $n$, i.e, $|p-q|$ for $p \cdot q = n$.}
    \label{fig: two_more_ansatz}
\end{figure}

\section{Factorization Records of different Methods}
Table \ref{tab:records} presents all factorization records we found for hybrid quantum-classical approaches. We do not claim completeness here.

\begin{table}[h!]
    \begin{tabular}{|l|l|l|l|l|}
    \hline
    \textbf{Number $n$} & \textbf{$N$ qubits} & \textbf{Algorithm} & \textbf{Hardware/Sim.} & \textbf{Year} \\ \hline
    15 & 7 & Shor & NMR & 2001 \cite{record1} \\ \hline
    15 & 4 & Shor & Photonic Qubits & 2007 \cite{Lanyon_2007} \\ \hline
    15 & 4 & Shor & Photonic Qubits & 2007 \cite{record25} \\ \hline
    21 & 3 & Adiabatic & NMR & 2008 \cite{record3} \\ \hline
    15 & 3 & Shor & Photonic & 2009 \cite{Politi_2009} \\ \hline
    143 & 4 & Adiabatic & NMR & 2011 \cite{record4} \\ \hline
    15 & 3 & Shor & Superconducting & 2012 \cite{record2} \\ \hline
    21 & 2 & Shor & Photonic & 2012 \cite{record28} \\ \hline
    11663 & 4 & Adiabatic & NMR & 2014 \cite{record6} \\ \hline
    291311 & 3 & Adiabatic & NMR & 2017 \cite{li2017highfidelity} \\ \hline
    4088459 & 4 & Grover & IBM & 2018 \cite{record10} \\ \hline
    291311 & 6 & VQF & IBM (SIM) & 2018 \cite{record11} \\ \hline
    376289 & 94 & Annealing & D-Wave (SIM) & 2018 \cite{record13} \\ \hline
    1005973 & 89 & Annealing & D-Wave (SIM)  & 2019 \cite{record9} \\ \hline
    1028171 & 88 & Annealing & D-Wave (SIM)  & 2020 \cite{record24} \\ \hline
    1099551473989 & 5 & VQF & IBM (SIM) & 2021 \cite{record16} \\ \hline
    21 & 5 & Shor & IBM & 2021 \cite{record26} \\ \hline
    1829 & 9 & QITE & IBM & 2021 \cite{record17} \\ \hline
    35 & 5 & Adiabatic & IBM & 2021 \cite{Saxena_2021} \\ \hline
    247 & 13 & VQF & IBM & 2022 \cite{record19} \\ \hline
    261980999226229 & 10 & Babai’s, Lattice, QAOA & IBM & 2022 \cite{record20} \\ \hline
    261980999226229 & 10 & DCQF & Trapped-ion & 2023 \cite{record18} \\ \hline
    102454763 & 26 & Annealing & D-Wave (SIM) & 2023 \cite{record22} \\ \hline
    1269636549803 & 4 & Grover & IBM & 2023 \cite{record27} \\ \hline
    1000070001221 & 12 & Annealing & D-Wave (SIM) & 2023 \cite{record22}\\ \hline
    \end{tabular}
    \label{tab:records}
\end{table}
\end{widetext}

\clearpage
\bibliography{references}

\begin{thebibliography}{49}%
\makeatletter
\providecommand \@ifxundefined [1]{%
 \@ifx{#1\undefined}
}%
\providecommand \@ifnum [1]{%
 \ifnum #1\expandafter \@firstoftwo
 \else \expandafter \@secondoftwo
 \fi
}%
\providecommand \@ifx [1]{%
 \ifx #1\expandafter \@firstoftwo
 \else \expandafter \@secondoftwo
 \fi
}%
\providecommand \natexlab [1]{#1}%
\providecommand \enquote  [1]{``#1''}%
\providecommand \bibnamefont  [1]{#1}%
\providecommand \bibfnamefont [1]{#1}%
\providecommand \citenamefont [1]{#1}%
\providecommand \href@noop [0]{\@secondoftwo}%
\providecommand \href [0]{\begingroup \@sanitize@url \@href}%
\providecommand \@href[1]{\@@startlink{#1}\@@href}%
\providecommand \@@href[1]{\endgroup#1\@@endlink}%
\providecommand \@sanitize@url [0]{\catcode `\\12\catcode `\$12\catcode `\&12\catcode `\#12\catcode `\^12\catcode `\_12\catcode `\%12\relax}%
\providecommand \@@startlink[1]{}%
\providecommand \@@endlink[0]{}%
\providecommand \url  [0]{\begingroup\@sanitize@url \@url }%
\providecommand \@url [1]{\endgroup\@href {#1}{\urlprefix }}%
\providecommand \urlprefix  [0]{URL }%
\providecommand \Eprint [0]{\href }%
\providecommand \doibase [0]{http://dx.doi.org/}%
\providecommand \selectlanguage [0]{\@gobble}%
\providecommand \bibinfo  [0]{\@secondoftwo}%
\providecommand \bibfield  [0]{\@secondoftwo}%
\providecommand \translation [1]{[#1]}%
\providecommand \BibitemOpen [0]{}%
\providecommand \bibitemStop [0]{}%
\providecommand \bibitemNoStop [0]{.\EOS\space}%
\providecommand \EOS [0]{\spacefactor3000\relax}%
\providecommand \BibitemShut  [1]{\csname bibitem#1\endcsname}%
\let\auto@bib@innerbib\@empty
\bibitem [{\citenamefont {Shor}(1994)}]{shor}%
  \BibitemOpen
  \bibfield  {author} {\bibinfo {author} {\bibfnamefont {P.W.}\ \bibnamefont {Shor}},\ }\bibfield  {title} {\enquote {\bibinfo {title} {Algorithms for quantum computation: discrete logarithms and factoring},}\ }in\ \href {\doibase 10.1109/SFCS.1994.365700} {\emph {\bibinfo {booktitle} {Proceedings 35th Annual Symposium on Foundations of Computer Science}}}\ (\bibinfo {year} {1994})\ pp.\ \bibinfo {pages} {124--134}\BibitemShut {NoStop}%
\bibitem [{\citenamefont {Skosana}\ and\ \citenamefont {Tame}(2021)}]{record26}%
  \BibitemOpen
  \bibfield  {author} {\bibinfo {author} {\bibfnamefont {Unathi}\ \bibnamefont {Skosana}}\ and\ \bibinfo {author} {\bibfnamefont {Mark}\ \bibnamefont {Tame}},\ }\bibfield  {title} {\enquote {\bibinfo {title} {Demonstration of shor’s factoring algorithm for n = 21 on ibm quantum processors},}\ }\href {\doibase 10.1038/s41598-021-95973-w} {\bibfield  {journal} {\bibinfo  {journal} {Scientific Reports}\ }\textbf {\bibinfo {volume} {11}} (\bibinfo {year} {2021}),\ 10.1038/s41598-021-95973-w}\BibitemShut {NoStop}%
\bibitem [{\citenamefont {Preskill}(2018)}]{nisq}%
  \BibitemOpen
  \bibfield  {author} {\bibinfo {author} {\bibfnamefont {John}\ \bibnamefont {Preskill}},\ }\bibfield  {title} {\enquote {\bibinfo {title} {Quantum computing in the nisq era and beyond},}\ }\href {\doibase 10.22331/q-2018-08-06-79} {\bibfield  {journal} {\bibinfo  {journal} {Quantum}\ }\textbf {\bibinfo {volume} {2}},\ \bibinfo {pages} {79} (\bibinfo {year} {2018})}\BibitemShut {NoStop}%
\bibitem [{\citenamefont {Nannicini}(2019)}]{PhysRevE99013304}%
  \BibitemOpen
  \bibfield  {author} {\bibinfo {author} {\bibfnamefont {Giacomo}\ \bibnamefont {Nannicini}},\ }\bibfield  {title} {\enquote {\bibinfo {title} {Performance of hybrid quantum-classical variational heuristics for combinatorial optimization},}\ }\href {\doibase 10.1103/PhysRevE.99.013304} {\bibfield  {journal} {\bibinfo  {journal} {Phys. Rev. E}\ }\textbf {\bibinfo {volume} {99}},\ \bibinfo {pages} {013304} (\bibinfo {year} {2019})}\BibitemShut {NoStop}%
\bibitem [{\citenamefont {Chai}\ \emph {et~al.}(2023)\citenamefont {Chai}, \citenamefont {Funcke}, \citenamefont {Hartung}, \citenamefont {Jansen}, \citenamefont {Kuehn}, \citenamefont {Stornati},\ and\ \citenamefont {Stollenwerk}}]{FGA_PRApllied}%
  \BibitemOpen
  \bibfield  {author} {\bibinfo {author} {\bibfnamefont {Yahui}\ \bibnamefont {Chai}}, \bibinfo {author} {\bibfnamefont {Lena}\ \bibnamefont {Funcke}}, \bibinfo {author} {\bibfnamefont {Tobias}\ \bibnamefont {Hartung}}, \bibinfo {author} {\bibfnamefont {Karl}\ \bibnamefont {Jansen}}, \bibinfo {author} {\bibfnamefont {Stefan}\ \bibnamefont {Kuehn}}, \bibinfo {author} {\bibfnamefont {Paolo}\ \bibnamefont {Stornati}}, \ and\ \bibinfo {author} {\bibfnamefont {Tobias}\ \bibnamefont {Stollenwerk}},\ }\bibfield  {title} {\enquote {\bibinfo {title} {Optimal flight-gate assignment on a digital quantum computer},}\ }\href {\doibase 10.1103/physrevapplied.20.064025} {\bibfield  {journal} {\bibinfo  {journal} {Physical Review Applied}\ }\textbf {\bibinfo {volume} {20}} (\bibinfo {year} {2023}),\ 10.1103/physrevapplied.20.064025}\BibitemShut {NoStop}%
\bibitem [{\citenamefont {Amaro}\ \emph {et~al.}(2022)\citenamefont {Amaro}, \citenamefont {Rosenkranz}, \citenamefont {Fitzpatrick}, \citenamefont {Hirano},\ and\ \citenamefont {Fiorentini}}]{Amaro_2022}%
  \BibitemOpen
  \bibfield  {author} {\bibinfo {author} {\bibfnamefont {David}\ \bibnamefont {Amaro}}, \bibinfo {author} {\bibfnamefont {Matthias}\ \bibnamefont {Rosenkranz}}, \bibinfo {author} {\bibfnamefont {Nathan}\ \bibnamefont {Fitzpatrick}}, \bibinfo {author} {\bibfnamefont {Koji}\ \bibnamefont {Hirano}}, \ and\ \bibinfo {author} {\bibfnamefont {Mattia}\ \bibnamefont {Fiorentini}},\ }\bibfield  {title} {\enquote {\bibinfo {title} {A case study of variational quantum algorithms for a job shop scheduling problem},}\ }\href {\doibase 10.1140/epjqt/s40507-022-00123-4} {\bibfield  {journal} {\bibinfo  {journal} {EPJ Quantum Technology}\ }\textbf {\bibinfo {volume} {9}} (\bibinfo {year} {2022}),\ 10.1140/epjqt/s40507-022-00123-4}\BibitemShut {NoStop}%
\bibitem [{\citenamefont {Peruzzo}\ \emph {et~al.}(2014)\citenamefont {Peruzzo}, \citenamefont {McClean}, \citenamefont {Shadbolt}, \citenamefont {Yung}, \citenamefont {Zhou}, \citenamefont {Love}, \citenamefont {Aspuru-Guzik},\ and\ \citenamefont {O'Brien}}]{vqe}%
  \BibitemOpen
  \bibfield  {author} {\bibinfo {author} {\bibfnamefont {Alberto}\ \bibnamefont {Peruzzo}}, \bibinfo {author} {\bibfnamefont {Jarrod}\ \bibnamefont {McClean}}, \bibinfo {author} {\bibfnamefont {Peter}\ \bibnamefont {Shadbolt}}, \bibinfo {author} {\bibfnamefont {Man-Hong}\ \bibnamefont {Yung}}, \bibinfo {author} {\bibfnamefont {Xiao-Qi}\ \bibnamefont {Zhou}}, \bibinfo {author} {\bibfnamefont {Peter~J.}\ \bibnamefont {Love}}, \bibinfo {author} {\bibfnamefont {Al{\'a}n}\ \bibnamefont {Aspuru-Guzik}}, \ and\ \bibinfo {author} {\bibfnamefont {Jeremy~L.}\ \bibnamefont {O'Brien}},\ }\href@noop {} {\enquote {\bibinfo {title} {A variational eigenvalue solver on a photonic quantum processor},}\ } (\bibinfo {year} {2014}),\ \Eprint {http://arxiv.org/abs/1304.3061} {arXiv:1304.3061 [quant-ph]} \BibitemShut {NoStop}%
\bibitem [{\citenamefont {Burges}(2002)}]{burges}%
  \BibitemOpen
  \bibfield  {author} {\bibinfo {author} {\bibfnamefont {Chris~J.C.}\ \bibnamefont {Burges}},\ }\href {https://www.microsoft.com/en-us/research/publication/factoring-as-optimization/} {\emph {\bibinfo {title} {Factoring as Optimization}}},\ \bibinfo {type} {Tech. Rep.}\ \bibinfo {number} {MSR-TR-2002-83}\ (\bibinfo {year} {2002})\BibitemShut {NoStop}%
\bibitem [{\citenamefont {Rockafellar}\ and\ \citenamefont {Uryasev}(2002)}]{rockafellar2000cvar}%
  \BibitemOpen
  \bibfield  {author} {\bibinfo {author} {\bibfnamefont {R.~Tyrrell}\ \bibnamefont {Rockafellar}}\ and\ \bibinfo {author} {\bibfnamefont {Stanislav}\ \bibnamefont {Uryasev}},\ }\bibfield  {title} {\enquote {\bibinfo {title} {Conditional value-at-risk for general loss distributions},}\ }\href@noop {} {\bibfield  {journal} {\bibinfo  {journal} {Journal of Banking \& Finance}\ }\textbf {\bibinfo {volume} {26}},\ \bibinfo {pages} {1443--1471} (\bibinfo {year} {2002})}\BibitemShut {NoStop}%
\bibitem [{\citenamefont {Barkoutsos}\ \emph {et~al.}(2020)\citenamefont {Barkoutsos}, \citenamefont {Nannicini}, \citenamefont {Robert}, \citenamefont {Tavernelli},\ and\ \citenamefont {Woerner}}]{Barkoutsos2020}%
  \BibitemOpen
  \bibfield  {author} {\bibinfo {author} {\bibfnamefont {Panagiotis~Kl}\ \bibnamefont {Barkoutsos}}, \bibinfo {author} {\bibfnamefont {Giacomo}\ \bibnamefont {Nannicini}}, \bibinfo {author} {\bibfnamefont {Anton}\ \bibnamefont {Robert}}, \bibinfo {author} {\bibfnamefont {Ivano}\ \bibnamefont {Tavernelli}}, \ and\ \bibinfo {author} {\bibfnamefont {Stefan}\ \bibnamefont {Woerner}},\ }\bibfield  {title} {\enquote {\bibinfo {title} {Improving {Variational} {Quantum} {Optimization} using {CVaR}},}\ }\href {\doibase 10.22331/q-2020-04-20-256} {\bibfield  {journal} {\bibinfo  {journal} {Quantum}\ }\textbf {\bibinfo {volume} {4}},\ \bibinfo {pages} {256} (\bibinfo {year} {2020})},\ \bibinfo {note} {arXiv:1907.04769 [quant-ph]}\BibitemShut {NoStop}%
\bibitem [{\citenamefont {Wang}\ \emph {et~al.}(2020)\citenamefont {Wang}, \citenamefont {Hu}, \citenamefont {Yao},\ and\ \citenamefont {Wang}}]{record24}%
  \BibitemOpen
  \bibfield  {author} {\bibinfo {author} {\bibfnamefont {Baonan}\ \bibnamefont {Wang}}, \bibinfo {author} {\bibfnamefont {Feng}\ \bibnamefont {Hu}}, \bibinfo {author} {\bibfnamefont {Haonan}\ \bibnamefont {Yao}}, \ and\ \bibinfo {author} {\bibfnamefont {Chao}\ \bibnamefont {Wang}},\ }\bibfield  {title} {\enquote {\bibinfo {title} {Prime factorization algorithm based on parameter optimization of ising model},}\ }\href {\doibase 10.1038/s41598-020-62802-5} {\bibfield  {journal} {\bibinfo  {journal} {Scientific Reports}\ }\textbf {\bibinfo {volume} {10}},\ \bibinfo {pages} {7106} (\bibinfo {year} {2020})}\BibitemShut {NoStop}%
\bibitem [{\citenamefont {Farhi}\ \emph {et~al.}(2001)\citenamefont {Farhi}, \citenamefont {Goldstone}, \citenamefont {Gutmann}, \citenamefont {Lapan}, \citenamefont {Lundgren},\ and\ \citenamefont {Preda}}]{Farhi_2001}%
  \BibitemOpen
  \bibfield  {author} {\bibinfo {author} {\bibfnamefont {Edward}\ \bibnamefont {Farhi}}, \bibinfo {author} {\bibfnamefont {Jeffrey}\ \bibnamefont {Goldstone}}, \bibinfo {author} {\bibfnamefont {Sam}\ \bibnamefont {Gutmann}}, \bibinfo {author} {\bibfnamefont {Joshua}\ \bibnamefont {Lapan}}, \bibinfo {author} {\bibfnamefont {Andrew}\ \bibnamefont {Lundgren}}, \ and\ \bibinfo {author} {\bibfnamefont {Daniel}\ \bibnamefont {Preda}},\ }\bibfield  {title} {\enquote {\bibinfo {title} {A quantum adiabatic evolution algorithm applied to random instances of an np-complete problem},}\ }\href {\doibase 10.1126/science.1057726} {\bibfield  {journal} {\bibinfo  {journal} {Science}\ }\textbf {\bibinfo {volume} {292}},\ \bibinfo {pages} {472–475} (\bibinfo {year} {2001})}\BibitemShut {NoStop}%
\bibitem [{\citenamefont {Born}\ and\ \citenamefont {Fock}(1928)}]{adiabatic}%
  \BibitemOpen
  \bibfield  {author} {\bibinfo {author} {\bibfnamefont {M.}~\bibnamefont {Born}}\ and\ \bibinfo {author} {\bibfnamefont {V.}~\bibnamefont {Fock}},\ }\bibfield  {title} {\enquote {\bibinfo {title} {Beweis des adiabatensatzes},}\ }\href@noop {} {\bibfield  {journal} {\bibinfo  {journal} {Z. Phys.}\ }\textbf {\bibinfo {volume} {51}},\ \bibinfo {pages} {165--180} (\bibinfo {year} {1928})}\BibitemShut {NoStop}%
\bibitem [{\citenamefont {Peng}\ \emph {et~al.}(2008)\citenamefont {Peng}, \citenamefont {Liao}, \citenamefont {Xu}, \citenamefont {Qin}, \citenamefont {Zhou}, \citenamefont {Suter},\ and\ \citenamefont {Du}}]{record3}%
  \BibitemOpen
  \bibfield  {author} {\bibinfo {author} {\bibfnamefont {Xinhua}\ \bibnamefont {Peng}}, \bibinfo {author} {\bibfnamefont {Zeyang}\ \bibnamefont {Liao}}, \bibinfo {author} {\bibfnamefont {Nanyang}\ \bibnamefont {Xu}}, \bibinfo {author} {\bibfnamefont {Gan}\ \bibnamefont {Qin}}, \bibinfo {author} {\bibfnamefont {Xianyi}\ \bibnamefont {Zhou}}, \bibinfo {author} {\bibfnamefont {Dieter}\ \bibnamefont {Suter}}, \ and\ \bibinfo {author} {\bibfnamefont {Jiangfeng}\ \bibnamefont {Du}},\ }\bibfield  {title} {\enquote {\bibinfo {title} {Quantum adiabatic algorithm for factorization and its experimental implementation},}\ }\href {\doibase 10.1103/physrevlett.101.220405} {\bibfield  {journal} {\bibinfo  {journal} {Physical Review Letters}\ }\textbf {\bibinfo {volume} {101}} (\bibinfo {year} {2008}),\ 10.1103/physrevlett.101.220405}\BibitemShut {NoStop}%
\bibitem [{\citenamefont {Li}\ \emph {et~al.}(2017)\citenamefont {Li}, \citenamefont {Dattani}, \citenamefont {Chen}, \citenamefont {Liu}, \citenamefont {Wang}, \citenamefont {Tanburn}, \citenamefont {Chen}, \citenamefont {Peng},\ and\ \citenamefont {Du}}]{li2017highfidelity}%
  \BibitemOpen
  \bibfield  {author} {\bibinfo {author} {\bibfnamefont {Zhaokai}\ \bibnamefont {Li}}, \bibinfo {author} {\bibfnamefont {Nikesh~S.}\ \bibnamefont {Dattani}}, \bibinfo {author} {\bibfnamefont {Xi}~\bibnamefont {Chen}}, \bibinfo {author} {\bibfnamefont {Xiaomei}\ \bibnamefont {Liu}}, \bibinfo {author} {\bibfnamefont {Hengyan}\ \bibnamefont {Wang}}, \bibinfo {author} {\bibfnamefont {Richard}\ \bibnamefont {Tanburn}}, \bibinfo {author} {\bibfnamefont {Hongwei}\ \bibnamefont {Chen}}, \bibinfo {author} {\bibfnamefont {Xinhua}\ \bibnamefont {Peng}}, \ and\ \bibinfo {author} {\bibfnamefont {Jiangfeng}\ \bibnamefont {Du}},\ }\href@noop {} {\enquote {\bibinfo {title} {High-fidelity adiabatic quantum computation using the intrinsic hamiltonian of a spin system: Application to the experimental factorization of 291311},}\ } (\bibinfo {year} {2017}),\ \Eprint {http://arxiv.org/abs/1706.08061} {arXiv:1706.08061 [quant-ph]} \BibitemShut {NoStop}%
\bibitem [{\citenamefont {Saxena}\ \emph {et~al.}(2021)\citenamefont {Saxena}, \citenamefont {Shukla},\ and\ \citenamefont {Pathak}}]{Saxena_2021}%
  \BibitemOpen
  \bibfield  {author} {\bibinfo {author} {\bibfnamefont {Ashwin}\ \bibnamefont {Saxena}}, \bibinfo {author} {\bibfnamefont {Abhishek}\ \bibnamefont {Shukla}}, \ and\ \bibinfo {author} {\bibfnamefont {Anirban}\ \bibnamefont {Pathak}},\ }\bibfield  {title} {\enquote {\bibinfo {title} {A hybrid scheme for prime factorization and its experimental implementation using ibm quantum processor},}\ }\href {\doibase 10.1007/s11128-021-03053-9} {\bibfield  {journal} {\bibinfo  {journal} {Quantum Information Processing}\ }\textbf {\bibinfo {volume} {20}} (\bibinfo {year} {2021}),\ 10.1007/s11128-021-03053-9}\BibitemShut {NoStop}%
\bibitem [{dwa()}]{dwave}%
  \BibitemOpen
  \href {https://docs.dwavesys.com/docs/latest/c_gs_2.html} {\enquote {\bibinfo {title} {What is quantum annealing? - d-wave system documentation},}\ }\bibinfo {note} {Accessed on: 5 February 2024}\BibitemShut {NoStop}%
\bibitem [{\citenamefont {Jiang}\ \emph {et~al.}(2018)\citenamefont {Jiang}, \citenamefont {Britt}, \citenamefont {McCaskey}, \citenamefont {Humble},\ and\ \citenamefont {Kais}}]{record13}%
  \BibitemOpen
  \bibfield  {author} {\bibinfo {author} {\bibfnamefont {Shuxian}\ \bibnamefont {Jiang}}, \bibinfo {author} {\bibfnamefont {Keith~A.}\ \bibnamefont {Britt}}, \bibinfo {author} {\bibfnamefont {Alexander~J.}\ \bibnamefont {McCaskey}}, \bibinfo {author} {\bibfnamefont {Travis~S.}\ \bibnamefont {Humble}}, \ and\ \bibinfo {author} {\bibfnamefont {Sabre}\ \bibnamefont {Kais}},\ }\href@noop {} {\enquote {\bibinfo {title} {Quantum annealing for prime factorization},}\ } (\bibinfo {year} {2018}),\ \Eprint {http://arxiv.org/abs/1804.02733} {arXiv:1804.02733 [quant-ph]} \BibitemShut {NoStop}%
\bibitem [{\citenamefont {Jun}\ and\ \citenamefont {Lee}(2023)}]{record22}%
  \BibitemOpen
  \bibfield  {author} {\bibinfo {author} {\bibfnamefont {Kyungtaek}\ \bibnamefont {Jun}}\ and\ \bibinfo {author} {\bibfnamefont {Hyunju}\ \bibnamefont {Lee}},\ }\bibfield  {title} {\enquote {\bibinfo {title} {Hubo and qubo models for prime factorization},}\ }\href {\doibase 10.1038/s41598-023-36813-x} {\bibfield  {journal} {\bibinfo  {journal} {Scientific Reports}\ }\textbf {\bibinfo {volume} {13}} (\bibinfo {year} {2023}),\ 10.1038/s41598-023-36813-x}\BibitemShut {NoStop}%
\bibitem [{\citenamefont {Anschuetz}\ \emph {et~al.}(2018)\citenamefont {Anschuetz}, \citenamefont {Olson}, \citenamefont {Aspuru-Guzik},\ and\ \citenamefont {Cao}}]{record11}%
  \BibitemOpen
  \bibfield  {author} {\bibinfo {author} {\bibfnamefont {Eric~R.}\ \bibnamefont {Anschuetz}}, \bibinfo {author} {\bibfnamefont {Jonathan~P.}\ \bibnamefont {Olson}}, \bibinfo {author} {\bibfnamefont {Alán}\ \bibnamefont {Aspuru-Guzik}}, \ and\ \bibinfo {author} {\bibfnamefont {Yudong}\ \bibnamefont {Cao}},\ }\href@noop {} {\enquote {\bibinfo {title} {Variational quantum factoring},}\ } (\bibinfo {year} {2018}),\ \Eprint {http://arxiv.org/abs/1808.08927} {arXiv:1808.08927 [quant-ph]} \BibitemShut {NoStop}%
\bibitem [{\citenamefont {Farhi}\ \emph {et~al.}(2014)\citenamefont {Farhi}, \citenamefont {Goldstone},\ and\ \citenamefont {Gutmann}}]{farhi2014quantum}%
  \BibitemOpen
  \bibfield  {author} {\bibinfo {author} {\bibfnamefont {Edward}\ \bibnamefont {Farhi}}, \bibinfo {author} {\bibfnamefont {Jeffrey}\ \bibnamefont {Goldstone}}, \ and\ \bibinfo {author} {\bibfnamefont {Sam}\ \bibnamefont {Gutmann}},\ }\href {https://arxiv.org/abs/1411.4028} {\enquote {\bibinfo {title} {A quantum approximate optimization algorithm},}\ } (\bibinfo {year} {2014}),\ \Eprint {http://arxiv.org/abs/1411.4028} {arXiv:1411.4028 [quant-ph]} \BibitemShut {NoStop}%
\bibitem [{\citenamefont {Karamlou}\ \emph {et~al.}(2021)\citenamefont {Karamlou}, \citenamefont {Simon}, \citenamefont {Katabarwa}, \citenamefont {Scholten}, \citenamefont {Peropadre},\ and\ \citenamefont {Cao}}]{record16}%
  \BibitemOpen
  \bibfield  {author} {\bibinfo {author} {\bibfnamefont {Amir~H.}\ \bibnamefont {Karamlou}}, \bibinfo {author} {\bibfnamefont {William~A.}\ \bibnamefont {Simon}}, \bibinfo {author} {\bibfnamefont {Amara}\ \bibnamefont {Katabarwa}}, \bibinfo {author} {\bibfnamefont {Travis~L.}\ \bibnamefont {Scholten}}, \bibinfo {author} {\bibfnamefont {Borja}\ \bibnamefont {Peropadre}}, \ and\ \bibinfo {author} {\bibfnamefont {Yudong}\ \bibnamefont {Cao}},\ }\href@noop {} {\enquote {\bibinfo {title} {Analyzing the performance of variational quantum factoring on a superconducting quantum processor},}\ } (\bibinfo {year} {2021}),\ \Eprint {http://arxiv.org/abs/2012.07825} {arXiv:2012.07825 [quant-ph]} \BibitemShut {NoStop}%
\bibitem [{\citenamefont {Phan}\ \emph {et~al.}(2022)\citenamefont {Phan}, \citenamefont {Pönni}, \citenamefont {Raasakka},\ and\ \citenamefont {Tittonen}}]{record19}%
  \BibitemOpen
  \bibfield  {author} {\bibinfo {author} {\bibfnamefont {Vivian}\ \bibnamefont {Phan}}, \bibinfo {author} {\bibfnamefont {Arttu}\ \bibnamefont {Pönni}}, \bibinfo {author} {\bibfnamefont {Matti}\ \bibnamefont {Raasakka}}, \ and\ \bibinfo {author} {\bibfnamefont {Ilkka}\ \bibnamefont {Tittonen}},\ }\href@noop {} {\enquote {\bibinfo {title} {On quantum factoring using noisy intermediate scale quantum computers},}\ } (\bibinfo {year} {2022}),\ \Eprint {http://arxiv.org/abs/2208.07085} {arXiv:2208.07085 [quant-ph]} \BibitemShut {NoStop}%
\bibitem [{\citenamefont {Motta}\ \emph {et~al.}(2019)\citenamefont {Motta}, \citenamefont {Sun}, \citenamefont {Tan}, \citenamefont {O’Rourke}, \citenamefont {Ye}, \citenamefont {Minnich}, \citenamefont {Brandão},\ and\ \citenamefont {Chan}}]{Motta_2019}%
  \BibitemOpen
  \bibfield  {author} {\bibinfo {author} {\bibfnamefont {Mario}\ \bibnamefont {Motta}}, \bibinfo {author} {\bibfnamefont {Chong}\ \bibnamefont {Sun}}, \bibinfo {author} {\bibfnamefont {Adrian T.~K.}\ \bibnamefont {Tan}}, \bibinfo {author} {\bibfnamefont {Matthew~J.}\ \bibnamefont {O’Rourke}}, \bibinfo {author} {\bibfnamefont {Erika}\ \bibnamefont {Ye}}, \bibinfo {author} {\bibfnamefont {Austin~J.}\ \bibnamefont {Minnich}}, \bibinfo {author} {\bibfnamefont {Fernando G. S.~L.}\ \bibnamefont {Brandão}}, \ and\ \bibinfo {author} {\bibfnamefont {Garnet Kin-Lic}\ \bibnamefont {Chan}},\ }\bibfield  {title} {\enquote {\bibinfo {title} {Determining eigenstates and thermal states on a quantum computer using quantum imaginary time evolution},}\ }\href {\doibase 10.1038/s41567-019-0704-4} {\bibfield  {journal} {\bibinfo  {journal} {Nature Physics}\ }\textbf {\bibinfo {volume} {16}},\ \bibinfo {pages} {205–210} (\bibinfo {year} {2019})}\BibitemShut {NoStop}%
\bibitem [{\citenamefont {McArdle}\ \emph {et~al.}(2019)\citenamefont {McArdle}, \citenamefont {Jones}, \citenamefont {Endo}, \citenamefont {Li}, \citenamefont {Benjamin},\ and\ \citenamefont {Yuan}}]{McArdle_2019}%
  \BibitemOpen
  \bibfield  {author} {\bibinfo {author} {\bibfnamefont {Sam}\ \bibnamefont {McArdle}}, \bibinfo {author} {\bibfnamefont {Tyson}\ \bibnamefont {Jones}}, \bibinfo {author} {\bibfnamefont {Suguru}\ \bibnamefont {Endo}}, \bibinfo {author} {\bibfnamefont {Ying}\ \bibnamefont {Li}}, \bibinfo {author} {\bibfnamefont {Simon~C.}\ \bibnamefont {Benjamin}}, \ and\ \bibinfo {author} {\bibfnamefont {Xiao}\ \bibnamefont {Yuan}},\ }\bibfield  {title} {\enquote {\bibinfo {title} {Variational ansatz-based quantum simulation of imaginary time evolution},}\ }\href {\doibase 10.1038/s41534-019-0187-2} {\bibfield  {journal} {\bibinfo  {journal} {npj Quantum Information}\ }\textbf {\bibinfo {volume} {5}} (\bibinfo {year} {2019}),\ 10.1038/s41534-019-0187-2}\BibitemShut {NoStop}%
\bibitem [{\citenamefont {Selvarajan}\ \emph {et~al.}(2021)\citenamefont {Selvarajan}, \citenamefont {Dixit}, \citenamefont {Cui}, \citenamefont {Humble},\ and\ \citenamefont {Kais}}]{record17}%
  \BibitemOpen
  \bibfield  {author} {\bibinfo {author} {\bibfnamefont {Raja}\ \bibnamefont {Selvarajan}}, \bibinfo {author} {\bibfnamefont {Vivek}\ \bibnamefont {Dixit}}, \bibinfo {author} {\bibfnamefont {Xingshan}\ \bibnamefont {Cui}}, \bibinfo {author} {\bibfnamefont {Travis~S.}\ \bibnamefont {Humble}}, \ and\ \bibinfo {author} {\bibfnamefont {Sabre}\ \bibnamefont {Kais}},\ }\href@noop {} {\enquote {\bibinfo {title} {Prime factorization using quantum variational imaginary time evolution},}\ } (\bibinfo {year} {2021}),\ \Eprint {http://arxiv.org/abs/2107.10196} {arXiv:2107.10196 [quant-ph]} \BibitemShut {NoStop}%
\bibitem [{\citenamefont {Dhaulakhandi}\ \emph {et~al.}(2023)\citenamefont {Dhaulakhandi}, \citenamefont {Behera},\ and\ \citenamefont {Seo}}]{record27}%
  \BibitemOpen
  \bibfield  {author} {\bibinfo {author} {\bibfnamefont {Ritu}\ \bibnamefont {Dhaulakhandi}}, \bibinfo {author} {\bibfnamefont {Bikash~K.}\ \bibnamefont {Behera}}, \ and\ \bibinfo {author} {\bibfnamefont {Felix~J.}\ \bibnamefont {Seo}},\ }\href@noop {} {\enquote {\bibinfo {title} {Factorization of large tetra and penta prime numbers on ibm quantum processor},}\ } (\bibinfo {year} {2023}),\ \Eprint {http://arxiv.org/abs/2304.04999} {arXiv:2304.04999 [quant-ph]} \BibitemShut {NoStop}%
\bibitem [{\citenamefont {Yan}\ \emph {et~al.}(2022)\citenamefont {Yan}, \citenamefont {Tan}, \citenamefont {Wei}, \citenamefont {Jiang}, \citenamefont {Wang}, \citenamefont {Wang}, \citenamefont {Luo}, \citenamefont {Duan}, \citenamefont {Liu}, \citenamefont {Shi}, \citenamefont {Fei}, \citenamefont {Meng}, \citenamefont {Han}, \citenamefont {Shan}, \citenamefont {Chen}, \citenamefont {Zhu}, \citenamefont {Zhang}, \citenamefont {Jin}, \citenamefont {Li}, \citenamefont {Song}, \citenamefont {Wang}, \citenamefont {Ma}, \citenamefont {Wang},\ and\ \citenamefont {Long}}]{record20}%
  \BibitemOpen
  \bibfield  {author} {\bibinfo {author} {\bibfnamefont {Bao}\ \bibnamefont {Yan}}, \bibinfo {author} {\bibfnamefont {Ziqi}\ \bibnamefont {Tan}}, \bibinfo {author} {\bibfnamefont {Shijie}\ \bibnamefont {Wei}}, \bibinfo {author} {\bibfnamefont {Haocong}\ \bibnamefont {Jiang}}, \bibinfo {author} {\bibfnamefont {Weilong}\ \bibnamefont {Wang}}, \bibinfo {author} {\bibfnamefont {Hong}\ \bibnamefont {Wang}}, \bibinfo {author} {\bibfnamefont {Lan}\ \bibnamefont {Luo}}, \bibinfo {author} {\bibfnamefont {Qianheng}\ \bibnamefont {Duan}}, \bibinfo {author} {\bibfnamefont {Yiting}\ \bibnamefont {Liu}}, \bibinfo {author} {\bibfnamefont {Wenhao}\ \bibnamefont {Shi}}, \bibinfo {author} {\bibfnamefont {Yangyang}\ \bibnamefont {Fei}}, \bibinfo {author} {\bibfnamefont {Xiangdong}\ \bibnamefont {Meng}}, \bibinfo {author} {\bibfnamefont {Yu}~\bibnamefont {Han}}, \bibinfo {author} {\bibfnamefont {Zheng}\ \bibnamefont {Shan}}, \bibinfo {author} {\bibfnamefont {Jiachen}\ \bibnamefont {Chen}}, \bibinfo {author} {\bibfnamefont
  {Xuhao}\ \bibnamefont {Zhu}}, \bibinfo {author} {\bibfnamefont {Chuanyu}\ \bibnamefont {Zhang}}, \bibinfo {author} {\bibfnamefont {Feitong}\ \bibnamefont {Jin}}, \bibinfo {author} {\bibfnamefont {Hekang}\ \bibnamefont {Li}}, \bibinfo {author} {\bibfnamefont {Chao}\ \bibnamefont {Song}}, \bibinfo {author} {\bibfnamefont {Zhen}\ \bibnamefont {Wang}}, \bibinfo {author} {\bibfnamefont {Zhi}\ \bibnamefont {Ma}}, \bibinfo {author} {\bibfnamefont {H.}~\bibnamefont {Wang}}, \ and\ \bibinfo {author} {\bibfnamefont {Gui-Lu}\ \bibnamefont {Long}},\ }\href@noop {} {\enquote {\bibinfo {title} {Factoring integers with sublinear resources on a superconducting quantum processor},}\ } (\bibinfo {year} {2022}),\ \Eprint {http://arxiv.org/abs/2212.12372} {arXiv:2212.12372 [quant-ph]} \BibitemShut {NoStop}%
\bibitem [{\citenamefont {Hegade}\ and\ \citenamefont {Solano}(2023)}]{record18}%
  \BibitemOpen
  \bibfield  {author} {\bibinfo {author} {\bibfnamefont {Narendra~N.}\ \bibnamefont {Hegade}}\ and\ \bibinfo {author} {\bibfnamefont {Enrique}\ \bibnamefont {Solano}},\ }\href@noop {} {\enquote {\bibinfo {title} {Digitized-counterdiabatic quantum factorization},}\ } (\bibinfo {year} {2023}),\ \Eprint {http://arxiv.org/abs/2301.11005} {arXiv:2301.11005 [quant-ph]} \BibitemShut {NoStop}%
\bibitem [{\citenamefont {Powell}(1994)}]{Powell1994}%
  \BibitemOpen
  \bibfield  {author} {\bibinfo {author} {\bibfnamefont {M.~J.~D.}\ \bibnamefont {Powell}},\ }\enquote {\bibinfo {title} {A direct search optimization method that models the objective and constraint functions by linear interpolation},}\ \ (\bibinfo  {publisher} {Springer Netherlands},\ \bibinfo {address} {Dordrecht},\ \bibinfo {year} {1994})\ pp.\ \bibinfo {pages} {51--67}\BibitemShut {NoStop}%
\bibitem [{\citenamefont {Ragonneau}\ and\ \citenamefont {Zhang}(2023)}]{ragonneau2023pdfo}%
  \BibitemOpen
  \bibfield  {author} {\bibinfo {author} {\bibfnamefont {Tom~M.}\ \bibnamefont {Ragonneau}}\ and\ \bibinfo {author} {\bibfnamefont {Zaikun}\ \bibnamefont {Zhang}},\ }\href@noop {} {\enquote {\bibinfo {title} {Pdfo: A cross-platform package for powell's derivative-free optimization solvers},}\ } (\bibinfo {year} {2023}),\ \Eprint {http://arxiv.org/abs/2302.13246} {arXiv:2302.13246 [math.OC]} \BibitemShut {NoStop}%
\bibitem [{\citenamefont {Pablo}\ \emph {et~al.}(2021)\citenamefont {Pablo}, \citenamefont {Diego},\ and\ \citenamefont {José}}]{D_ez_Valle_2021}%
  \BibitemOpen
  \bibfield  {author} {\bibinfo {author} {\bibfnamefont {Díez-Vallej}\ \bibnamefont {Pablo}}, \bibinfo {author} {\bibfnamefont {Porras}\ \bibnamefont {Diego}}, \ and\ \bibinfo {author} {\bibfnamefont {García-Ripoll~Juan}\ \bibnamefont {José}},\ }\bibfield  {title} {\enquote {\bibinfo {title} {Quantum variational optimization: The role of entanglement and problem hardness},}\ }\href {\doibase 10.1103/physreva.104.062426} {\bibfield  {journal} {\bibinfo  {journal} {Physical Review A}\ }\textbf {\bibinfo {volume} {104}} (\bibinfo {year} {2021}),\ 10.1103/physreva.104.062426}\BibitemShut {NoStop}%
\bibitem [{\citenamefont {Ragonneau}(2023)}]{ragonneau2023modelbased}%
  \BibitemOpen
  \bibfield  {author} {\bibinfo {author} {\bibfnamefont {Tom~M.}\ \bibnamefont {Ragonneau}},\ }\href@noop {} {\enquote {\bibinfo {title} {Model-based derivative-free optimization methods and software},}\ } (\bibinfo {year} {2023}),\ \Eprint {http://arxiv.org/abs/2210.12018} {arXiv:2210.12018 [math.OC]} \BibitemShut {NoStop}%
\bibitem [{sta()}]{statevector}%
  \BibitemOpen
  \href {www.qiskit.org/ecosystem/aer/stubs/qiskit_aer.AerSimulator.html} {\enquote {\bibinfo {title} {Aersimulator},}\ }\bibinfo {note} {Accessed on: 14 November 2023}\BibitemShut {NoStop}%
\bibitem [{qis()}]{qiskit}%
  \BibitemOpen
  \href {https://qiskit.org/} {\enquote {\bibinfo {title} {Qiskit is the open-source toolkit for useful quantum},}\ }\bibinfo {note} {Accessed on: 14 November 2023}\BibitemShut {NoStop}%
\bibitem [{reg()}]{regression}%
  \BibitemOpen
  \href {https://www.statsmodels.org/stable/regression.html} {\enquote {\bibinfo {title} {Linear regression},}\ }\bibinfo {note} {Accessed on: 12 February 2024}\BibitemShut {NoStop}%
\bibitem [{\citenamefont {Nakanishi}\ \emph {et~al.}(2020)\citenamefont {Nakanishi}, \citenamefont {Fujii},\ and\ \citenamefont {Todo}}]{nft}%
  \BibitemOpen
  \bibfield  {author} {\bibinfo {author} {\bibfnamefont {Ken~M.}\ \bibnamefont {Nakanishi}}, \bibinfo {author} {\bibfnamefont {Keisuke}\ \bibnamefont {Fujii}}, \ and\ \bibinfo {author} {\bibfnamefont {Synge}\ \bibnamefont {Todo}},\ }\bibfield  {title} {\enquote {\bibinfo {title} {Sequential minimal optimization for quantum-classical hybrid algorithms},}\ }\href {\doibase 10.1103/physrevresearch.2.043158} {\bibfield  {journal} {\bibinfo  {journal} {Physical Review Research}\ }\textbf {\bibinfo {volume} {2}} (\bibinfo {year} {2020}),\ 10.1103/physrevresearch.2.043158}\BibitemShut {NoStop}%
\bibitem [{\citenamefont {Cerezo}\ \emph {et~al.}(2021)\citenamefont {Cerezo}, \citenamefont {Sone}, \citenamefont {Volkoff}, \citenamefont {Cincio},\ and\ \citenamefont {Coles}}]{Cerezo_2021}%
  \BibitemOpen
  \bibfield  {author} {\bibinfo {author} {\bibfnamefont {M.}~\bibnamefont {Cerezo}}, \bibinfo {author} {\bibfnamefont {Akira}\ \bibnamefont {Sone}}, \bibinfo {author} {\bibfnamefont {Tyler}\ \bibnamefont {Volkoff}}, \bibinfo {author} {\bibfnamefont {Lukasz}\ \bibnamefont {Cincio}}, \ and\ \bibinfo {author} {\bibfnamefont {Patrick~J.}\ \bibnamefont {Coles}},\ }\bibfield  {title} {\enquote {\bibinfo {title} {Cost function dependent barren plateaus in shallow parametrized quantum circuits},}\ }\href {\doibase 10.1038/s41467-021-21728-w} {\bibfield  {journal} {\bibinfo  {journal} {Nature Communications}\ }\textbf {\bibinfo {volume} {12}} (\bibinfo {year} {2021}),\ 10.1038/s41467-021-21728-w}\BibitemShut {NoStop}%
\bibitem [{\citenamefont {Chai}\ \emph {et~al.}(2024)\citenamefont {Chai}, \citenamefont {Jansen}, \citenamefont {Kühn}, \citenamefont {Schwägerl},\ and\ \citenamefont {Stollenwerk}}]{chai2024structureinspiredansatzwarmstart}%
  \BibitemOpen
  \bibfield  {author} {\bibinfo {author} {\bibfnamefont {Yahui}\ \bibnamefont {Chai}}, \bibinfo {author} {\bibfnamefont {Karl}\ \bibnamefont {Jansen}}, \bibinfo {author} {\bibfnamefont {Stefan}\ \bibnamefont {Kühn}}, \bibinfo {author} {\bibfnamefont {Tim}\ \bibnamefont {Schwägerl}}, \ and\ \bibinfo {author} {\bibfnamefont {Tobias}\ \bibnamefont {Stollenwerk}},\ }\href {https://arxiv.org/abs/2407.02569} {\enquote {\bibinfo {title} {Structure-inspired ansatz and warm start of variational quantum algorithms for quadratic unconstrained binary optimization problems},}\ } (\bibinfo {year} {2024}),\ \Eprint {http://arxiv.org/abs/2407.02569} {arXiv:2407.02569 [quant-ph]} \BibitemShut {NoStop}%
\bibitem [{\citenamefont {Vandersypen}\ \emph {et~al.}(2001)\citenamefont {Vandersypen}, \citenamefont {Steffen}, \citenamefont {Breyta}, \citenamefont {Yannoni}, \citenamefont {Sherwood},\ and\ \citenamefont {Chuang}}]{record1}%
  \BibitemOpen
  \bibfield  {author} {\bibinfo {author} {\bibfnamefont {Lieven M.~K.}\ \bibnamefont {Vandersypen}}, \bibinfo {author} {\bibfnamefont {Matthias}\ \bibnamefont {Steffen}}, \bibinfo {author} {\bibfnamefont {Gregory}\ \bibnamefont {Breyta}}, \bibinfo {author} {\bibfnamefont {Costantino~S.}\ \bibnamefont {Yannoni}}, \bibinfo {author} {\bibfnamefont {Mark~H.}\ \bibnamefont {Sherwood}}, \ and\ \bibinfo {author} {\bibfnamefont {Isaac~L.}\ \bibnamefont {Chuang}},\ }\bibfield  {title} {\enquote {\bibinfo {title} {Experimental realization of shor’s quantum factoring algorithm using nuclear magnetic resonance},}\ }\href {\doibase 10.1038/414883a} {\bibfield  {journal} {\bibinfo  {journal} {Nature}\ }\textbf {\bibinfo {volume} {414}},\ \bibinfo {pages} {883–887} (\bibinfo {year} {2001})}\BibitemShut {NoStop}%
\bibitem [{\citenamefont {P.}\ \emph {et~al.}(2007)\citenamefont {P.}, \citenamefont {J.}, \citenamefont {K.}, \citenamefont {M.}, \citenamefont {V.}, \citenamefont {A.},\ and\ \citenamefont {White}}]{Lanyon_2007}%
  \BibitemOpen
  \bibfield  {author} {\bibinfo {author} {\bibfnamefont {Lanyon~B.}\ \bibnamefont {P.}}, \bibinfo {author} {\bibfnamefont {Weinhold~T.}\ \bibnamefont {J.}}, \bibinfo {author} {\bibfnamefont {Langford~N.}\ \bibnamefont {K.}}, \bibinfo {author} {\bibfnamefont {Barbieri}\ \bibnamefont {M.}}, \bibinfo {author} {\bibfnamefont {James D.~F.}\ \bibnamefont {V.}}, \bibinfo {author} {\bibfnamefont {Gilchrist}\ \bibnamefont {A.}}, \ and\ \bibinfo {author} {\bibnamefont {White}},\ }\bibfield  {title} {\enquote {\bibinfo {title} {Experimental demonstration of a compiled version of shor’s algorithm with quantum entanglement},}\ }\href {\doibase 10.1103/physrevlett.99.250505} {\bibfield  {journal} {\bibinfo  {journal} {Physical Review Letters}\ }\textbf {\bibinfo {volume} {99}} (\bibinfo {year} {2007}),\ 10.1103/physrevlett.99.250505}\BibitemShut {NoStop}%
\bibitem [{\citenamefont {Chao-Yang}\ \emph {et~al.}(2007)\citenamefont {Chao-Yang}, \citenamefont {Browne}, \citenamefont {E.}, \citenamefont {Yang}, \citenamefont {Tao},\ and\ \citenamefont {Jian-Wei}}]{record25}%
  \BibitemOpen
  \bibfield  {author} {\bibinfo {author} {\bibfnamefont {Lu}~\bibnamefont {Chao-Yang}}, \bibinfo {author} {\bibnamefont {Browne}}, \bibinfo {author} {\bibfnamefont {Daniel}\ \bibnamefont {E.}}, \bibinfo {author} {\bibnamefont {Yang}}, \bibinfo {author} {\bibnamefont {Tao}}, \ and\ \bibinfo {author} {\bibfnamefont {Pan}\ \bibnamefont {Jian-Wei}},\ }\bibfield  {title} {\enquote {\bibinfo {title} {Demonstration of a compiled version of shor’s quantum factoring algorithm using photonic qubits},}\ }\href {\doibase 10.1103/physrevlett.99.250504} {\bibfield  {journal} {\bibinfo  {journal} {Physical Review Letters}\ }\textbf {\bibinfo {volume} {99}} (\bibinfo {year} {2007}),\ 10.1103/physrevlett.99.250504}\BibitemShut {NoStop}%
\bibitem [{\citenamefont {Alberto}\ \emph {et~al.}(2009)\citenamefont {Alberto}, \citenamefont {F.},\ and\ \citenamefont {L.}}]{Politi_2009}%
  \BibitemOpen
  \bibfield  {author} {\bibinfo {author} {\bibfnamefont {Politi}\ \bibnamefont {Alberto}}, \bibinfo {author} {\bibfnamefont {Matthews Jonathan~C.}\ \bibnamefont {F.}}, \ and\ \bibinfo {author} {\bibfnamefont {O’Brien~Jeremy}\ \bibnamefont {L.}},\ }\bibfield  {title} {\enquote {\bibinfo {title} {Shor’s quantum factoring algorithm on a photonic chip},}\ }\href {\doibase 10.1126/science.1173731} {\bibfield  {journal} {\bibinfo  {journal} {Science}\ }\textbf {\bibinfo {volume} {325}},\ \bibinfo {pages} {1221–1221} (\bibinfo {year} {2009})}\BibitemShut {NoStop}%
\bibitem [{\citenamefont {Xu}\ \emph {et~al.}(2012)\citenamefont {Xu}, \citenamefont {Zhu}, \citenamefont {Lu}, \citenamefont {Zhou}, \citenamefont {Peng},\ and\ \citenamefont {Du}}]{record4}%
  \BibitemOpen
  \bibfield  {author} {\bibinfo {author} {\bibfnamefont {Nanyang}\ \bibnamefont {Xu}}, \bibinfo {author} {\bibfnamefont {Jing}\ \bibnamefont {Zhu}}, \bibinfo {author} {\bibfnamefont {Dawei}\ \bibnamefont {Lu}}, \bibinfo {author} {\bibfnamefont {Xianyi}\ \bibnamefont {Zhou}}, \bibinfo {author} {\bibfnamefont {Xinhua}\ \bibnamefont {Peng}}, \ and\ \bibinfo {author} {\bibfnamefont {Jiangfeng}\ \bibnamefont {Du}},\ }\bibfield  {title} {\enquote {\bibinfo {title} {Erratum: Quantum factorization of 143 on a dipolar-coupling nuclear magnetic resonance system},}\ }\href {\doibase 10.1103/physrevlett.109.269902} {\bibfield  {journal} {\bibinfo  {journal} {Physical Review Letters}\ }\textbf {\bibinfo {volume} {109}} (\bibinfo {year} {2012}),\ 10.1103/physrevlett.109.269902}\BibitemShut {NoStop}%
\bibitem [{\citenamefont {Lucero}\ \emph {et~al.}(2012)\citenamefont {Lucero}, \citenamefont {Barends}, \citenamefont {Chen}, \citenamefont {Kelly}, \citenamefont {Mariantoni}, \citenamefont {Megrant}, \citenamefont {O’Malley}, \citenamefont {Sank}, \citenamefont {Vainsencher}, \citenamefont {Wenner}, \citenamefont {White}, \citenamefont {Yin}, \citenamefont {Cleland},\ and\ \citenamefont {Martinis}}]{record2}%
  \BibitemOpen
  \bibfield  {author} {\bibinfo {author} {\bibfnamefont {Erik}\ \bibnamefont {Lucero}}, \bibinfo {author} {\bibfnamefont {R.}~\bibnamefont {Barends}}, \bibinfo {author} {\bibfnamefont {Y.}~\bibnamefont {Chen}}, \bibinfo {author} {\bibfnamefont {J.}~\bibnamefont {Kelly}}, \bibinfo {author} {\bibfnamefont {M.}~\bibnamefont {Mariantoni}}, \bibinfo {author} {\bibfnamefont {A.}~\bibnamefont {Megrant}}, \bibinfo {author} {\bibfnamefont {P.}~\bibnamefont {O’Malley}}, \bibinfo {author} {\bibfnamefont {D.}~\bibnamefont {Sank}}, \bibinfo {author} {\bibfnamefont {A.}~\bibnamefont {Vainsencher}}, \bibinfo {author} {\bibfnamefont {J.}~\bibnamefont {Wenner}}, \bibinfo {author} {\bibfnamefont {T.}~\bibnamefont {White}}, \bibinfo {author} {\bibfnamefont {Y.}~\bibnamefont {Yin}}, \bibinfo {author} {\bibfnamefont {A.~N.}\ \bibnamefont {Cleland}}, \ and\ \bibinfo {author} {\bibfnamefont {John~M.}\ \bibnamefont {Martinis}},\ }\bibfield  {title} {\enquote {\bibinfo {title} {Computing prime factors with a josephson phase qubit
  quantum processor},}\ }\href {\doibase 10.1038/nphys2385} {\bibfield  {journal} {\bibinfo  {journal} {Nature Physics}\ }\textbf {\bibinfo {volume} {8}},\ \bibinfo {pages} {719–723} (\bibinfo {year} {2012})}\BibitemShut {NoStop}%
\bibitem [{\citenamefont {Martín-López}\ \emph {et~al.}(2012)\citenamefont {Martín-López}, \citenamefont {Laing}, \citenamefont {Lawson}, \citenamefont {Alvarez}, \citenamefont {Zhou},\ and\ \citenamefont {O’Brien}}]{record28}%
  \BibitemOpen
  \bibfield  {author} {\bibinfo {author} {\bibfnamefont {Enrique}\ \bibnamefont {Martín-López}}, \bibinfo {author} {\bibfnamefont {Anthony}\ \bibnamefont {Laing}}, \bibinfo {author} {\bibfnamefont {Thomas}\ \bibnamefont {Lawson}}, \bibinfo {author} {\bibfnamefont {Roberto}\ \bibnamefont {Alvarez}}, \bibinfo {author} {\bibfnamefont {Xiao-Qi}\ \bibnamefont {Zhou}}, \ and\ \bibinfo {author} {\bibfnamefont {Jeremy~L.}\ \bibnamefont {O’Brien}},\ }\bibfield  {title} {\enquote {\bibinfo {title} {Experimental realization of shor’s quantum factoring algorithm using qubit recycling},}\ }\href {\doibase 10.1038/nphoton.2012.259} {\bibfield  {journal} {\bibinfo  {journal} {Nature Photonics}\ }\textbf {\bibinfo {volume} {6}},\ \bibinfo {pages} {773–776} (\bibinfo {year} {2012})}\BibitemShut {NoStop}%
\bibitem [{\citenamefont {Dattani}\ and\ \citenamefont {Bryans}(2014)}]{record6}%
  \BibitemOpen
  \bibfield  {author} {\bibinfo {author} {\bibfnamefont {Nikesh~S.}\ \bibnamefont {Dattani}}\ and\ \bibinfo {author} {\bibfnamefont {Nathaniel}\ \bibnamefont {Bryans}},\ }\href@noop {} {\enquote {\bibinfo {title} {Quantum factorization of 56153 with only 4 qubits},}\ } (\bibinfo {year} {2014}),\ \Eprint {http://arxiv.org/abs/1411.6758} {arXiv:1411.6758 [quant-ph]} \BibitemShut {NoStop}%
\bibitem [{\citenamefont {Dash}\ \emph {et~al.}(2018)\citenamefont {Dash}, \citenamefont {Sarmah}, \citenamefont {Behera},\ and\ \citenamefont {Panigrahi}}]{record10}%
  \BibitemOpen
  \bibfield  {author} {\bibinfo {author} {\bibfnamefont {Avinash}\ \bibnamefont {Dash}}, \bibinfo {author} {\bibfnamefont {Deepankar}\ \bibnamefont {Sarmah}}, \bibinfo {author} {\bibfnamefont {Bikash~K.}\ \bibnamefont {Behera}}, \ and\ \bibinfo {author} {\bibfnamefont {Prasanta~K.}\ \bibnamefont {Panigrahi}},\ }\href@noop {} {\enquote {\bibinfo {title} {Exact search algorithm to factorize large biprimes and a triprime on ibm quantum computer},}\ } (\bibinfo {year} {2018}),\ \Eprint {http://arxiv.org/abs/1805.10478} {arXiv:1805.10478 [quant-ph]} \BibitemShut {NoStop}%
\bibitem [{\citenamefont {Peng}\ \emph {et~al.}(2019)\citenamefont {Peng}, \citenamefont {Wang}, \citenamefont {Hu}, \citenamefont {Wang}, \citenamefont {Fang}, \citenamefont {Chen},\ and\ \citenamefont {Wang}}]{record9}%
  \BibitemOpen
  \bibfield  {author} {\bibinfo {author} {\bibfnamefont {WangChun}\ \bibnamefont {Peng}}, \bibinfo {author} {\bibfnamefont {BaoNan}\ \bibnamefont {Wang}}, \bibinfo {author} {\bibfnamefont {Feng}\ \bibnamefont {Hu}}, \bibinfo {author} {\bibfnamefont {YunJiang}\ \bibnamefont {Wang}}, \bibinfo {author} {\bibfnamefont {XianJin}\ \bibnamefont {Fang}}, \bibinfo {author} {\bibfnamefont {XingYuan}\ \bibnamefont {Chen}}, \ and\ \bibinfo {author} {\bibfnamefont {Chao}\ \bibnamefont {Wang}},\ }\bibfield  {title} {\enquote {\bibinfo {title} {Factoring larger integers with fewer qubits via quantum annealing with optimized parameters},}\ }\href {\doibase 10.1007/s11433-018-9307-1} {\bibfield  {journal} {\bibinfo  {journal} {Sci. China Phys. Mech. Astron}\ }\textbf {\bibinfo {volume} {62}} (\bibinfo {year} {2019}),\ 10.1007/s11433-018-9307-1}\BibitemShut {NoStop}%
\end{thebibliography}%
\end{document}